\documentclass[lettersize,journal]{IEEEtran}
\usepackage{algorithmic}
\usepackage{algorithm}
\usepackage{array}
\usepackage{textcomp}
\usepackage{stfloats}
\usepackage{url}
\usepackage{verbatim}
\usepackage{graphicx}
\usepackage{cite}
\usepackage{amsmath,amssymb,amsfonts}
\usepackage{float}
\usepackage{textcomp}
\usepackage{xcolor}
\usepackage{subfigure}
\usepackage{url}
\usepackage{makecell}
\usepackage{color}
\usepackage{verbatim}
\usepackage{hyperref}
\usepackage{multicol}
\usepackage{fancyhdr}
\def\BibTeX{{\rm B\kern-.05em{\sc i\kern-.025em b}\kern-.08em
    T\kern-.1667em\lower.7ex\hbox{E}\kern-.125emX}}

\begin{document}

\pagestyle{fancy}

\fancyhead[C]{This paper appears in IEEE Transactions on Wireless Communications (TWC), 2023.}

\title{User-centric Heterogeneous-action Deep Reinforcement Learning for Virtual {Reality in the Metaverse over Wireless Networks}}

\author{\IEEEauthorblockN{Wenhan Yu,~\IEEEmembership{Student Member,~IEEE, }Terence Jie Chua, and \\Jun Zhao,~\IEEEmembership{Member,~IEEE}}\vspace{-20pt} 
\thanks{The authors are all with Nanyang Technological University, Singapore. Corresponding author: Jun Zhao, Email: JunZHAO@ntu.edu.sg.
}
\thanks{A 6-page short version containing partial results is accepted to the 2023 IEEE International Conference on Communications (ICC)~\cite{ICC}.} 
}
\maketitle
\thispagestyle{fancy}
\begin{abstract}
The Metaverse emerging as maturing technologies are empowering the different facets. Virtual Reality (VR) technologies serve as the backbone of the virtual universe within the Metaverse to offer a highly immersive user experience. As mobility is emphasized in the Metaverse context, VR devices reduce their weights at the sacrifice of local computation abilities. In this paper, for a system consisting of a Metaverse server and multiple VR users, we consider two cases of (i) the server generating frames and transmitting them to users, and (ii) users generating frames locally and thus consuming device energy. As Metaverse emphasizes on the accessibility for all users anywhere and anytime, the users can have totally different characteristics, devices and demands. In this paper, the channel access arrangement (including the decisions on frame generation location), and transmission powers for the downlink communications from the server to the users are jointly optimized by our proposed user-centric Deep Reinforcement Learning (DRL) algorithm, namely User-centric Critic with Heterogenous Actors (UCHA). Comprehensive experiments demonstrate that our UCHA algorithm leads to remarkable results under various requirements and constraints.

\end{abstract}


\section{Introduction}
\subsection{Background}
The recent futuristic notion of Metaverse is an extension, a complete simulation, and a mirror of the real world, which is empowered by the maturing high-performance Extended Reality (XR) technologies. Among them, Virtual Reality (VR) provides users with a fully immersive and digital world, which has been used in many fields such as entertainment, socialization, and industry~\cite{metaverse, mobileVR}. To alleviate the obtrusive sense of restriction for users when moving around wearing VR devices, the mobility of these devices is essential. Feasible solutions are applied by using wireless connections and decreasing the weights at the sacrifice of local computation abilities. Thus, even state-of-art VR devices (e.g., HTC Vive~\cite{VRsurvey}) do not have the sufficient local computing power to support high-resolution and high-frame-rate applications. In this event, transferring some VR frame generations to a remote server is necessary. Furthermore, under the context of the Metaverse, the boundary between virtual and physical environments becomes more and more blurred, which brings more frequent demands of accessing the virtual world from users. As the Metaverse emphasizes building the bridge between the virtual and real worlds and supporting user inputs anywhere and anytime, it is necessary to consider how to allocate the limited resources to a wide variety of VR users (VUs) with distinct characteristics and demands.

\subsection{Challenges and motivations}
We first explain the challenges with motivations for this work from the following aspects.

\textbf{User-centric features and user-diverse problems.} 
The main reason for its popularity can be contributed to the user-centric services, which serve as the fundamental basis of the next-generation Internet~\cite{metaverse}. The Metaverse is a perceived virtual universe and ideally, allows people to access it anywhere and anytime, with any purpose. Compared to the traditional network structure, the user-centric features also pose a user-diverse problem: the wide variety of use cases (applications) and user-inherent characteristics (e.g., device types, battery power) induce highly different demands for frames per second (FPS), resolutions, etc. Further, the much larger size of transmitted data in VR can pose a challenge for devices with lower capabilities. These devices may struggle to complete tasks without sufficient support from remote computing resources. Therefore, the first and foremost challenge is how to achieve efficient utilization of those network resources in the multi-user scenario where each user has a different purpose of use and requirements. This propels us to seek a more user-centric and oriented solution to handle widely different users. 

\textbf{Wireless communication for VR.} VR is a key feature of an immersive Metaverse socialization experience. Compared to traditional two-dimensional images, generating $360^{\circ}$ panoramic images for the VR experience is computationally intensive. However, as the mobility of VR devices is attached with great importance under the Metaverse context, manufacturers have to lessen the weight at the cost of local computation capability. As a consequence, the existing VR devices lack the local computation ability of high-resolution and frame-rate applications. A feasible solution to powering an immersive socialization experience on VR devices is to make the Metaverse Server help the application frame generation, and send generated frames  to VUs. However, with the frequent demand and high congestion degrees for network resources, it is necessary to lighten the network burden by allocating some devices with higher local computing power to do local generation sometimes. Therefore, this paper considers two cases: (i) server generation, and (ii) local generation.

\textbf{Joint optimization in wireless communication.} In many scenarios, there are more than one important objectives that need to be optimized. In the wireless scenario where two cases of server and local generation are taken into account, two parts are detrimental to data transfer efficiency: (i) The channel access arrangement for VUs (including being assigned with no channel and doing local generation), and (ii) transmission power allocation for VUs. Based on this, this paper focuses on jointly optimizing these variables in the wireless downlink transmission scenario, and considering the VUs' diverse characteristics.

\subsection{Related work and our novelty} 
The related work is separated into multiple aspects according to the challenges and motivations since we design our work and make contributions from these multiple domains. The references and our novelties compared to them are expounded in the following.

\textbf{User-centric Metaverse.}
For the time being, people are already very aware of the user-centric particularity in the Metaverse. Lee \emph{et al.}~\cite{metaverse} claim that the Metaverse is user-centric by design, and it will rely on pervasive network access. Consequently, users with diverse purposes, devices, and demands can access the universe anytime and anywhere. Du \emph{et al.}~\cite{usercentricdu} emphasized the user-centric demands in the Metaverse and proposed an attention-aware network resource allocation considering the diversities among users' interests and applications. Both papers above came up with attractive and fresh concepts and discussed the potential challenges and future directions. However, none of them has ever studied a specific scenario with problems and designed novel algorithms to tackle them. In this paper, we design a concrete user-diverse problem scenario and invent a user-centric DRL structure correspondingly.

\textbf{VR over wireless communication.} In recent years, VR services over wireless communication have been thoroughly studied in many previous works. Yang \emph{et al.}~\cite{pengyang} investigated the problem of providing ultra-reliable and power-efficient VR strategies for wireless mobile VUs by Deep Reinforcement Learning (DRL) algorithms. In order to fit the discrete action space required in their DRL algorithm, they quantize the continuous actions into discrete actions. Xiao \emph{et al.}~\cite{chenxuVR} studied the predictive VR video delivering by optimizing the bitrate with DRL methods. Some other works has also demonstrated the excellent performance of DRL methods over wireless communications as its ability to explore and exploit in self-defined environments~\cite{chenxuwireless, chenxuwireless2, FLRL}. However, none of the previous works considered the varying purpose of use and requirements between different VUs, and there are no existing works that have designed a \textit{user-centric} and \textit{user-oriented} DRL method as our proposed solutions. 

\textbf{Joint optimization in wireless communications with DRL methods.}
Some remarkable works have investigated using DRL methods to solve joint optimization problems~\cite{MALS,asyRL}. For instance, Guo \emph{et al.}~\cite{JO1} solved the handover control and power allocation joint problem using Multi-agent Proximal Policy Optimization (MAPPO) and obtained satisfactory results. As mixed continuous-discrete actions can lead to problems of directly embedding the MAPPO algorithm, they use discrete power requirement factors instead of a continuous power allocation to simplify the problem. Thus, the problem they aimed to address does not have heterogeneous actions (both discrete and continuous actions), hence they can directly use MAPPO. He \emph{et al.}~\cite{JO2} researched the joint optimization problems on channel access and power resource arrangements by using DRL methods to determine the channel assignment and conducting traditional optimization methods for power allocation under the channel information. However, none of these works considers a user-diverse scenario or problems with heterogeneous actions. Compared to them, we propose a novel Multi-Agent Deep Reinforcement Learning (MADRL) structure, which is equipped with a user-centric view and able to handle interactive and heterogeneous actions.

\subsection{Methodology and Contributions} 
This paper proposes a novel multi-user VR model in a downlink Non-Orthogonal Multiple Access (NOMA) system. Specifically, we optimize the channel access arrangement and downlink power allocation jointly, taking the diversities between VUs into consideration. We designed a novel MADRL algorithm, User-centric Critic with Heterogenous Actors (UCHA), which considers the varying purpose of use and requirements of the users and handles the heterogeneous action spaces. In terms of the backbone of our algorithm, we re-design the widely-used Proximal Policy Optimization (PPO) algorithm~\cite{PPO} with a reward decomposition structure in Critic, and two assymetric Actors. 

Our contributions are as follows:

\begin{itemize}
\item \emph{Formulating user-centric VR in the Metaverse:} We study the user-centric Metaverse over the wireless network, designing a multi-user VR scenario where a Metaverse Server assists users in generating reality-assisted virtual environments.
\item \emph{Heterogeneous Actors for inseparable optimization variables:} We created two asymmetric Actors interacting with each other to handle the inseparable and discrete-continuous mixed optimization objectives. Specifically, Actor one is for the channel access arrangement, and Actor two is for the power allocation based on the solution by Actor one.
\item \emph{User-centric Critic for user-diverse scenario:} We crafted a novel user-centric Critic equipped with a more user-specific architecture, in which we decompose the reward into multiple VUs and evaluate the value for each VU. To the best of our knowledge, we are the first to embed the hybrid reward architecture in multi-agent reinforcement learning, and use it to solve communication problems.
\item \emph{Novel and comprehensive simulation:} We conduct comprehensive experiments and design novel metrics to evaluate our proposed solution. The experimental results indicate that UCHA achieves the fastest convergence speed and attains the highest rewards among all baseline models. UCHA's allocations are more user-specific to individual users and appear to be more reasonable, as they fulfill different requirements of the users.
\end{itemize}

\subsection{Organization} 

The rest of the paper is organized as follows. Section~\ref{sec:models} introduces our system model. Then Sections~\ref{sec:RLenv} and \ref{sec:algorithms} propose our deep reinforcement learning setting and algorithm. In Section~\ref{sec:experiment}, extensive experiments are performed, and various methods are compared to show the prowess of our strategy. Section~\ref{conclude} concludes the paper.

A 6-page short version is accepted by the 2023 IEEE International Conference on Communications (ICC)~\cite{ICC}. In that conference version, we compare the user-centric structure in Critic with the normal Critic structure, and demonstrate its remarkable performance and ability to accelerates the convergence speed. Besides, we also demonstrate that using this structure in PPO is much superior to Hybrid Reward DQN~\cite{HRA}. However, in that version, we do not consider anything about the resolution, nor the multi-agent structure for further optimizing the transmission power, and these are all highlights of this journal version. Furthermore, this paper also designs new algorithms, metrics, and evaluation methods, compared to the conference version.

\section{System Model}
\label{sec:models}

Consider a multi-user wireless downlink transmission in an indoor environment, in which $T$ frames are generated by the Metaverse server and sent to $N$ VR Users (VUs) in one second. To ensure a smooth experience, we apply the clock signal from the server for synchronization, and use a slotted time structure, where one second is divided into $T$ time slots (steps), and the duration of each slot is $\iota=\frac{1}{T}$. In each slot, a high-resolution 3D scene frame for each user is generated and sent to $N$ VUs $\mathcal{N}=\{1,2,...,N\}$ via a set of channels $\mathcal{M}=\{1,2,...,M\}$. These VUs have distinct characteristics (e.g., local computation capability) and different requirements (e.g., Frames Per Second (FPS)). This FPS requirement is decided by the specific applications. Assume that one VU is playing VR video games, while another VU is having a virtual meeting. The FPS requirement for the former should be evidently much higher than the latter. Each user can accept VR frame rates as low as a minimum tolerable frame per second (FPS) $\tau_{n,F}$, which is the number of successfully received frames in one second (i.e., $T$ time slots). Our objective is to obtain channel access and downlink power arrangements for each VU.

\subsection{Channel allocation and frame generation}

In terms of channel allocation, we define an $N \times T$ matrix $\boldsymbol{Z}$ such that the element in its $n$th row and $t$th column is $z_n^t$, for $n \in \{1,2,\ldots,N\}$ and $t \in \{1,2,\ldots,T\}$, indicating that the channel allocation for VU $n$ at $t$ is $z_n^t$. In other words, $\boldsymbol{Z}$ denotes the selection of downlink channel arrangement. Specifically, in the studied system of one Metaverse server and $N$ VUs, our channel allocation for each VU $n \in \{1,2\ldots,N\}$ includes two cases:
\begin{itemize}
\item \textbf{Case 1: Server-generated frame.} If VU $n$ is assigned for a channel $m$ at time step $t$ (i.e., $z_n^t=m, m\in\mathcal{M}$), the Metaverse server selects the $m$th channel for downlink communication with VU $n$ to deliver the frame that the server generates for VU $n$. In this case, if the sum delay (for generation and transmission) of each frame exceeds slot duration $\iota$, this frame is deemed as a failure.
\item \textbf{Case 2: VU-generated frame.} If VU $n$ is assigned with no channel (i.e.,$z_n^t=0$), it generates the frame locally with a lower computation capability, and at the expense of energy consumption without communicating with the server. In this case, each VU assigned to do local generation will generate the frame of the highest in-time-processable resolution (i.e., can be generated in $\iota$). If the generated resolution is below the minimum acceptable resolution, this frame is deemed as a failure. 
\end{itemize}

Thus, the channel allocation in this paper also includes the decisions on whether the frames for the VUs are generated by the server or the VUs (for simplicity, we sometimes just say channel allocation without mentioning decisions of \mbox{frame-generation} locations since the former includes the latter). In other words, $z_n^t=m,m\in\mathcal{M}$ indicates VU $n$ is arranged to channel $m$ at step $t$, and $z_n^t=0$ means user $n$ needs to generate locally. Next, we explain the two cases.

\begin{figure*}[t] 
\centering
\includegraphics[width=0.78\linewidth]{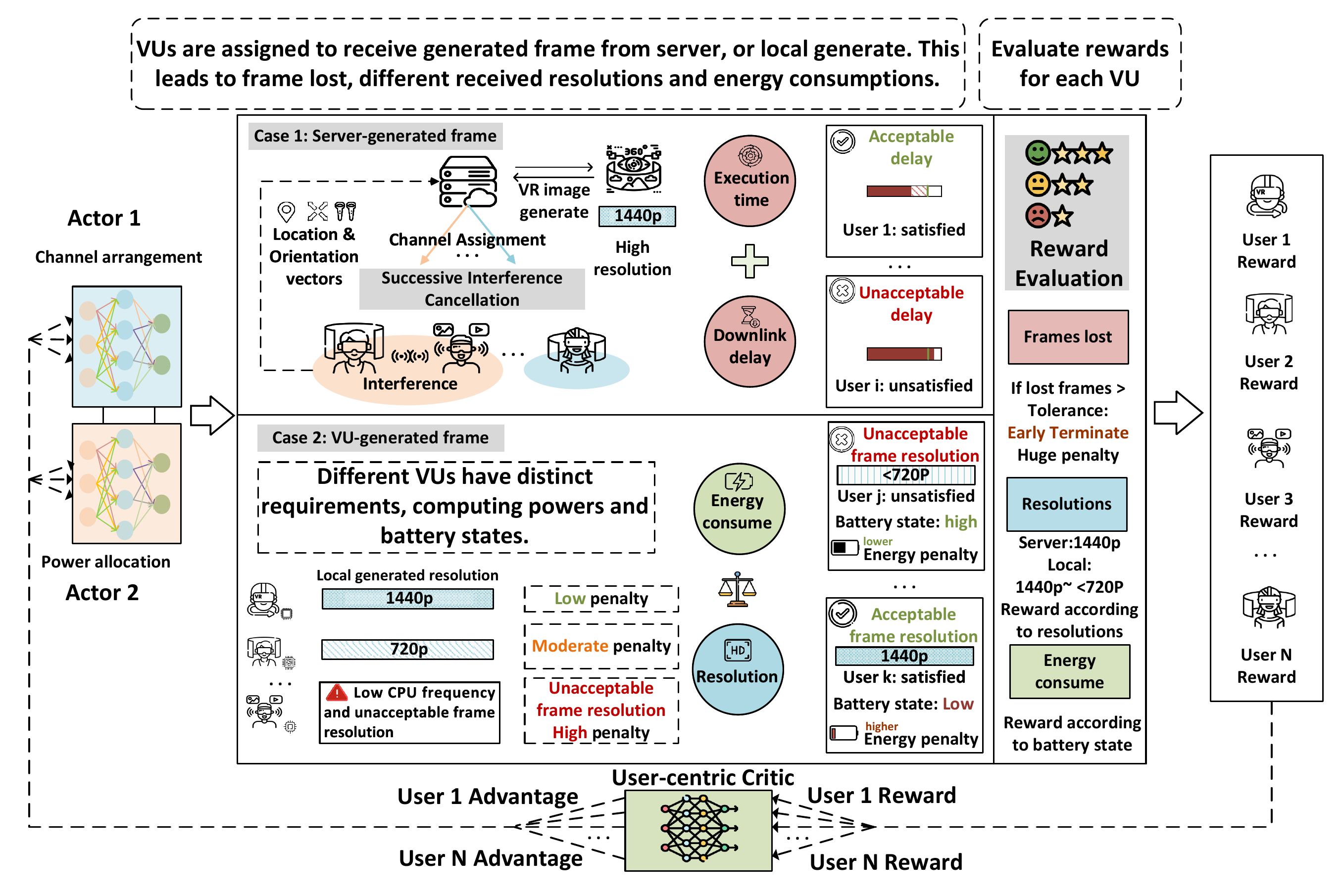}
\caption{System model. This figure illustrates a single time slot execution, where the frame generation contains two cases, server-generation and local-generation. Then, different metrics are evaluated and rewards are given accordingly. }
\label{fig:model}
\end{figure*}

\subsection{The case of the frame being generated and sent by server}\label{model:off-Model}
In each time step, the server will manage the downlink channels $\mathcal{M}$ of all VUs $\mathcal{N}$, and it will subsequently allocate the downlink transmission powers $\boldsymbol{p}^t$ with this Channel State Information (CSI). Here we define $\boldsymbol{P}:=[\boldsymbol{p}^1, \boldsymbol{p}^2, \ldots, \boldsymbol{p}^T]$ as the downlink power, where $\boldsymbol{p}^t:=[p_1^t, p_2^t, \ldots, p_N^t]$, $p_n^t$ is the power for VU $n$ at $t$, and enforce $\sum_{n\in\mathcal{N}}p_n^t\leq p_{max}$. As the total delay in this case and the resolutions are important metrics, the achievable rate and resolutions are discussed in the following:

\subsubsection{Achievable rate from the server to each VU}

We adopt the Non-Orthogonal Multiple Access (NOMA) system as this work's propagation model since it allows multiple users to share the same frequency band, which increases the capacity of the network~\cite{NOMA}. In our NOMA system, several VUs can be multiplexed on one channel by superposition coding, and each VU exploits successive interference cancellation (SIC) at its receiver. The decoding process follows the approach described by Dai~\textit{et~al.}~\cite{NOMA}. Specifically, with $\boldsymbol{z}^t$ denoting $[z_1^t,z_2^t,\ldots,z_N^t]$, then among the $N$ VUs, we let $\mathcal{N}_m^t(\boldsymbol{z}^t)$ be the set of $N_m^t(\boldsymbol{z}^t)$ users which receive VR frames from the channel $m$ at time step $t$. Formally, we have
\begin{align}
 \mathcal{N}_m^t(\boldsymbol{z}^t):= \{{n}\in\mathcal{N}~|~z_n^t=m\}, \text{ and } N_m^t(\boldsymbol{z}^t) = |\mathcal{N}_m^t(\boldsymbol{z}^t)|.  \label{eqNm}
\end{align}
The part ``$(\boldsymbol{z}^t)$'' means that $\mathcal{N}_m^t(\boldsymbol{z}^t)$ and $N_m^t(\boldsymbol{z}^t)$ are both functions $\boldsymbol{z}^t$. Based on~\cite{NOMA}, we order $N_m^t(\boldsymbol{z}^t)$ VUs of $\mathcal{N}_m^t(\boldsymbol{z}^t)$ such that channel-to-noise ratios between them and the server on channel $m$ are in a decreasing order; formally, recalling that  all $N$ VUs are indexed by $1,2,\ldots,N$, suppose that the above ordering for $N_m^t(\boldsymbol{z}^t)$ VUs of $\mathcal{N}_m^t(\boldsymbol{z}^t)$ produces VU indices $u_1, u_2, \ldots, u_{N_m^t(\boldsymbol{z}^t)}$; i.e., with $h_{i,m}^t$ denoting the channel attenuation between the server and VU $i$ over channel $m$ at time step $t$, and $(\sigma_{i,m}^t)^2$ denoting the power spectral density of additive white Gaussian noise at VU $i$ over channel $m$ at time step $t$, we have the following\footnote{We write $(\sigma_{i,m}^t)^2$ here for generality. In our experiments of Section~\ref{sec-Experiments}, $(\sigma_{i,m}^t)^2$ is the same $\sigma^2$ for any $i,m,t$.}: define $u_1, u_2, \ldots, u_{N_m^t(\boldsymbol{z}^t)}$ such that they together form the set $\mathcal{N}_m^t(\boldsymbol{z}^t)$ of Eq.~(\ref{eqNm}), and
\begin{align}
\frac{|h^t_{u_1,m}|^2}{(\sigma^t_{u_1,m})^2 }\geq \frac{|h^t_{u_2,m}|^2}{(\sigma^t_{u_2,m})^2 }\geq \ldots \geq \frac{|h^t_{u_{N_m^t(\boldsymbol{z}^t)},m}|^2}{(\sigma^t_{u_{N_m^t(\boldsymbol{z}^t)},m})^2 } 
\label{defineuseq}
\end{align}
Note that if two VUs with indices $u$ and $u'$ result in the same channel-to-noise ratios, their ordering can be arbitrary and this paper places the larger index ahead of the lower index; i.e., $u$ is before (resp., after) $u'$ if $u$ is greater (resp., smaller) than $u'$.  Following the same rationality as~\cite{NOMA}, after decoding via SIC, the interferences to VU $n$ satisfying $z_n^t=m$ come from signals sent from server that are intended for users whose positions are before VU $n$ in the sequence $u_1, u_2, \ldots, u_{N_m^t(\boldsymbol{z}^t)}$; formally, supposing $u_{\nu}$ is $n$ (such $\nu$ exists since $n\in\mathcal{N}_m^t(\boldsymbol{z}^t)$ follows from $z_n^t=m$), those users are $u_1, u_2, \ldots, u_{\nu -1}$ and the sum of the interferences to VU $n$ (i.e., $u_{\nu}$) are given by $\sum_{j=1}^{\nu -1} p_{u_j}^t |h_{n,m}^t|^2$, after we define $p_i^t$ as the transmit power used by server for transmitting the signal intended for VU $i$ at time step $t$. 

Then the achievable rate of VU $n$ over its assigned channel $z_n^t=m$  is
\begin{align}
   \hspace{-2pt}{r}_{n}^{t}(\boldsymbol{z}^t,\hspace{-1pt}\boldsymbol{p}^t)\hspace{-1pt}=\hspace{-1pt}W_m\log\hspace{-2pt}\left(\hspace{-2pt}1\hspace{-1pt}+\hspace{-1pt}\frac{{p}_{n}^{t}|h_{n,m}^{t}|^2}{\sum_{j=1}^{\nu -1} p_{u_j}^t |h_{n,m}^t|^2\hspace{-1pt}+\hspace{-1pt}W_m(\sigma_{n,m}^t)^2}\hspace{-2pt}\right)\hspace{-3pt},\label{eq:dlrate} 
\end{align}
for $\nu$ satisfying $u_{\nu}=n$ after defining related notations via~(\ref{eqNm}) and~(\ref{defineuseq}), where $W_m$ denotes the bandwidth of channel $m$,

\subsubsection{Resolution of server-generated frame}

We denote $D_{n,o}^{t}$ ($n \in \mathcal{N}$) as the transmission data size of the VR frame at time step $t$ that needs to be executed and transmitted by server to user $n$, $\mathcal{G}=\{G_{1}, ..., G_{J}\}$ as the frame sizes with different graphic resolutions in descending order (e.g., $G_1$ is 1440p, and $G_2$ is 1080p), where $G_{1}$ and $G_{J}$ are the lowest and highest acceptable resolutions, respectively. For convenience, we also define $G_{0} = +\infty$ and $G_{J+1}=0$. ${Res}_n^t$ is the received frame resolution of VU $n$ at $t$ (${Res}_n^t \in \mathcal{G}$). Assume that the server-generated frames are all in the highest resolution (which is $G_{1}$), and the transmission data size from server is
\begin{align}
    D_{n}^t = \frac{{Res}_n^t}{{Com}_{n}^{t}}.
\end{align}
In parallel to the fast-developing VR devices, video compression technologies are being developed as well. VR compression leverages the likeness of different images from different cameras and uses advanced slicing and tiling techniques~\cite{compression}. The compression ratio is not always a constant but variable to different VR image qualities and data sizes. To better fit the real situation, we use the $Com_n^t$ as the varying compression ratios of this frame of VU $n$ at time step $t$.

Accordingly, the delay $d_{n}^{t}(\boldsymbol{z}^t,\boldsymbol{p}^t)$ of each frame in time step $t$ is divided into (1) execution time and (2) downlink transmission time:
\begin{align}
    d_{n}^{t}(\boldsymbol{z}^t,\boldsymbol{p}^t) = \frac{D_n^t\times c_n^t}{f_v} + \frac{D_n^t}{{r}_{n}^{t}(\boldsymbol{z}^t,\boldsymbol{p}^t)}, \label{eq:offdelay}
\end{align}
where $f_v$ is the computation capability of the server (i.e., cycles per second), and $c_n^t$ is the required number of cycles per bit of this frame~\cite{COsettings}. 

\subsection{The case of the frame being generated by VUs locally}\label{model:local-Model}
When VU is not allocated a channel, it needs to generate the VR frames locally at the expense of \textbf{energy consumption} and \textbf{resolution degeneration} according to its local computing capability (CPU frequency). Let $f_n$ be the computation capability of VU $n$ , and it varies across VUs. Adopting the model from~\cite{energy}, the energy per cycle can be expressed as $e_{n,cyc} = \eta f_n^2$. Therefore, the energy consumption overhead of local computing can be derived as:
\begin{align}
    e_{n,l}^t = 
    \begin{cases}
        \mu_n \times D_n^t \times c_n^t \times e_{n,cyc}, &z_n^t \neq 0.\\
        0, &z_n^t =0.
    \end{cases}
     \label{eq:energy}
\end{align}
Inherently, $\mu_n$ is the battery weighting parameter of energy for VU $n$. The battery state of each VU can be different, then, we assume that $\mu_n$ is closer to $0$ with a higher battery. 

In terms of the resolution, if VU $n$ is doing local generation, the resolution of the current frame will degenerate to the highest resolution that VU $n$ can process with tolerable delay. Thus, the frame resolution of VU $n$ at $t$ is formulated as:
\begin{align}
    Res_{n}^t = 
    \begin{cases}
        G_{1}, &z_n^t \neq 0. \\
        G_{J_{n}^t}, &z_n^t = 0.
    \end{cases}
\end{align}
where
\begin{align}
    &G_{J_{n}^t}/{Com}_n^t \leq f_n \times \iota \leq G_{(J_{n}^t-1)}/{Com}_n^t,~\text{and}~~ 1 \leq J_{n}^t \leq J. \nonumber
\end{align}
Here, $J_{n}^t$ is the rank of the highest available resolution of VU $n$ among all resolution $\mathcal{D}$, and $J$ is the number of resolution types. $\iota$ is the duration of each time step ($\iota = \frac{1}{T}$), and $f_n\times \iota$ is the maximum datasize can be locally generated by VU $n$ in one step. The overall system model is shown in Fig.~\ref{fig:model}.

\subsection{Problem formulation}
Different users have different purposes for use (video games, group chat, etc.). Therefore, they also have varying expectations of satisfactory FPS $\tau_{n,F}$. For each successive frame, the total delay   the tolerable threshold (occurs in ``Case 1'') and the insufficient resolution ($Res_{n}^t < G_{J}$, occurs in ``Case 2'') lead to a frame failure. We set a frame success flag $I_n^t$ for VU $n$ at step $t$ as: 
\begin{align}
    I_n^t = 
    \begin{cases}
        0, &\text{if}~z_n^t \in \{1,2,\ldots,M\} \text{ and }d_{n}^{t}(\boldsymbol{z}^t,\boldsymbol{p}^t) > \iota. \\
        0, &\text{if}~z_n^t =0 \text{ and }Res_{n}^{t} < G_{J}. \\
        1, &\text{else}.
    \end{cases}
\end{align}
Our goal is to decide channel allocation and downlink power allocation for the transmission of $T$ frames. The objectives can be divided into 1) fulfill the FPS requirements of different users as possible and minimizing the total transmission failures, 2) optimize VUs' device energy usage regarding their battery state, and 3) increase VUs' received frames resolutions.

We define an indicator function $\chi[A]$ which takes $1$ if event $A$ occurs and takes $0$ otherwise. From the above discussion, our optimization problem is
\begin{align}
&\max\limits_{\boldsymbol{Z},\boldsymbol{P}}
\hspace{-2pt}\left \{ \hspace{-2pt}\omega_1 \hspace{-2pt}\left ( \hspace{-2pt} \min\limits_{n\in\mathcal{N}} \hspace{-2pt}\left[ \hspace{-1pt} \sum_{t=1}^T I_n^t\hspace{-2pt}-\hspace{-2pt}\tau_{n,F}\right] \hspace{-2pt} \right) \hspace{-2pt} + \hspace{-2pt} \frac{1}{N} \hspace{-2pt}\sum_{n=1}^N\sum_{t=1}^T[\omega_3 {Res}_n^t \hspace{-2pt} - \hspace{-2pt}\omega_2 e_{n,l}^t] \hspace{-2pt} \right \} \label{eq:objectivefunction}\\
&s.t.~C1:z_n^t\in\{0,1,...,M\},~\forall{n}\in\mathcal{N},\forall{t}\in\{1,2,\ldots,T\},\\
&~~~~~C2:\sum_{n\in \mathcal{N}} \chi[z_n^t \neq 0] p_n^t \leq p_{max}, \forall{t}\in\{1,2,\ldots,T\}.
\end{align}

The $\omega_1,\omega_2, \omega_3$ are the weighting parameters of frame failures, local device energy consumption, and VU devices received frames resolution. In practice, these parameters will be reflected in the reward setting~\ref{sec:RLenv}. The first part of the objective function (i.e., $\min\limits_{n\in\mathcal{N}} [(\sum_{t=1}^T I_n^t)-\tau_{n,F}]$) is to make every VU fulfill their FPS requirements as possible, and the second part is to minimize the local energy usage and maximize the frame resolutions. Constraint $C1$ is our integer optimization variable which denotes the computing method and channel assignment for each user at every time step. Constraint $C2$ is the limit of the downlink transmission power for all VUs.


\subsection{The execution flow of the system}

Here we describe the execution flow of the system given a solution to the optimization problem. The key idea of the two cases design (server or local generation) is to utilize the limited network resources and let some VUs spare the resources by doing local generation at each time. Thus, this paper uses a slotted time structure by applying the clock signal from the server for synchronization. Specifically, in each time slot $t$, one frame of each VU will be executed. Considering the very short duration $\iota$ of each time slot, we assume that the channel attenuation $h_{n,m}^t$ varies across different slots but remains the same in one slot. At the beginning of each slot $t$ (frame), the server collects the time-varying information of each VU (e.g., channel attenuation, compression rate), and sends the decisions on channel access (local or server generation, if server generation, which channel) to all VUs through a dedicated channel. Then, VUs assigned to no channel will locally generate the frame immediately, and the others will wait for the server to transmit the frames to them.

\subsection{Motivation of using MADRL in the optimization problem}

The optimization variables in the formulated problem, computation method choice, channel arrangement, and power allocation make the formulated problem highly coupled with inseparable mixed-integer non-linear programming (MINLP) optimization problems, where the discrete variable $\boldsymbol{Z}$ and continuous variable $\boldsymbol{P}$ are inseparable, which is NP-hard~\cite{NPhard}. Moreover, this formulated problem is \textbf{time-sequential}, where the number of variables increases with $T$. Thus, the traditional optimization methods are unsuitable for our proposed problem due to the daunting computational complexity. Also, as the problem contains too many random variables, \textit{model-based reinforcement learning (RL)} approaches that require transition probabilities are infeasible techniques to tackle our proposed problem. \textit{Heuristic search} methods can possibly be a solution to sequential problems. However, it doesn't change the approximation of the policy, while naively making improved action selections given the current value function. Besides, a huge tree of possible continuations is usually considered when using this method~\cite {RLintro}, which further makes it impractical to apply heuristic search in such a complicated scenario with a huge dimension of decisions. Therefore, it is highly necessary to design a comprehensive \mbox{model-free} Multi-Agent Deep Reinforcement Learning (MADRL) method to tackle the problem with heterogeneous optimization variables and distinct requirements from all VUs.

\section{Environment in Our Multi-Agent Deep Reinforcement Learning (MADRL)}
\label{sec:RLenv}
To tackle a problem with DRL method, designing a comprehensive reinforcement learning environment based on the formulated problem is the first and foremost step. For a reinforcement learning environment, the most important components are (1) State: the key factors for an agent to make a decision. (2) Action: the operation decided by an agent to interact with the environment. (3) Reward: the feedback for Agent to evaluate the action under this state. Thus, we expound on these three components next.

\subsection{State}
In the DRL environment, weeding out less relevant and less time-varying variables is essential. Therefore, we set two agents: $Agent_1$ for optimizing the channel arrangement, and $Agent_2$ for allocating the downlink transmission power. 

\subsubsection{$Agent_1$ State $s^t_1$} We included the following attributes into the state: (1) Each VU's frame size: $D_n^t={Res}_n^t/{Com}_n^t$. (2) Each VU's remaining tolerable frame transmission failure count. (3) The channel attenuation of each VU: $h_{n,m}^t$. (4) The remaining number of time slots: $(T-t)$. (5) Gap from requirement to each VU: $\tau_{n,F}-\sum_{i=1}^tI_n^i$.

\subsubsection{$Agent_2$ State $s^t_2$} Only after obtaining the CSI, the power allocation can be finished based on it. As a result, the \textbf{action} of $Agent_1$ is significant to the decision of $Agent_2$. Besides the action, the other important attributes in $s^t_1$ should be considered by $Agent_2$ as well. Therefore, we use the concatenation as the state: $s_2^t=\{a_1^t;s_1^t\}$.

\subsection{Action}
The appropriate action settings that are directly related to the optimization variables are critical for finding a good solution. In this environment, the action $a_1^t$ and action $a_2^t$ are respectively the $Agent_1$ and $Agent_2$ actions, explained below.  \\
\begin{figure}[t]
    \centering
    \setlength{\abovecaptionskip}{-0.1cm}
    \includegraphics[width=1\linewidth]{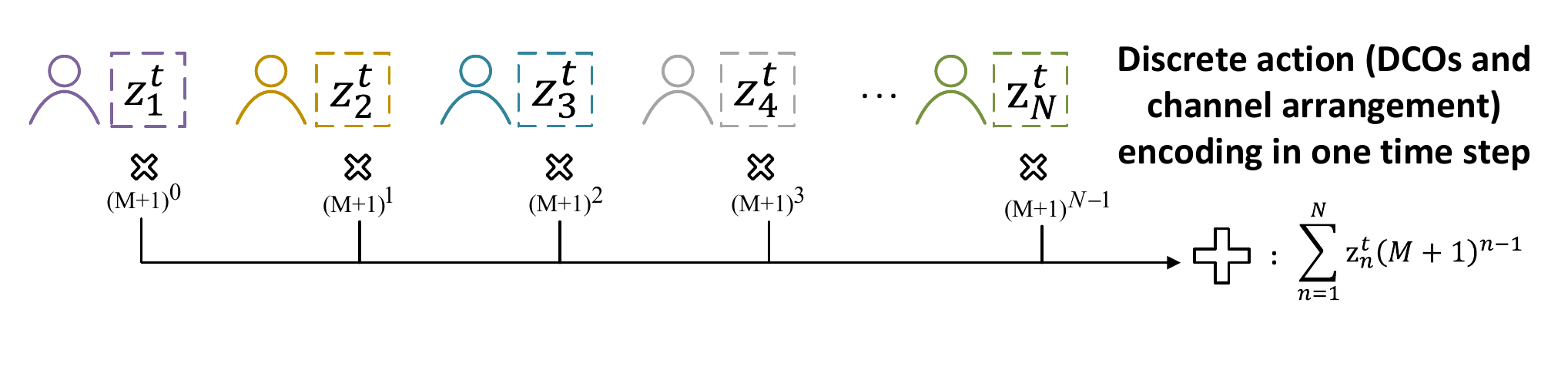}
    \caption{UL action encoding method. The action $a_1^t = \sum_{n=1}^{N}z_n^t(M+1)^{n-1}$.}
    \label{fig:actionencode}
\end{figure}
\subsubsection{$Agent_1$ Action $a_1^t$}
The discrete action of $Agent_1$ is the channel allocation: $\boldsymbol{z}^t = \{z_1^t, z_2^t,..., z_N^t\}$. In a DRL environment, the discrete actions are actually discrete numbers (indicators), then, we need to give consecutive discrete indicators to each action. As $\boldsymbol{z}^t$ contains $N$ elements ($N$ VUs), and each element has $M+1$ possible values ($M$ channels and plus 1 for decision on the cases), we use the $N$-bit-\mbox{$(M+1)$}-number code and decode it to decimal indicator as shown in Fig.~\ref{fig:actionencode}:
\begin{align}
    a_1^t = \sum_{n=1}^{N}z_n^t(M+1)^{n-1},~\forall t\in \{1,\cdots,T\}.
\end{align}
\subsubsection{$Agent_2$ Action $a_2^t$}
The continuous action of $Agent_2$ should be the downlink transmission power $\{p_1^t,p_2^t,...,p_N^t\}$. However, it is impractical for the RL agent to allocate power for all VU within the sum power constraint. Therefore, we add a softmax layer to the $Agent_2$. Accordingly, the $a_2^t$ turns into the portions of the $p_{max}$. We use $P_n^t$ denotes the portion allocated to $n$ at $t$:
\begin{align}
    a_2^t = \{P_1^t, P_2^t,...,P_N^t\},~\forall t\in \{1,\cdots,T\}.
\end{align}

\subsection{Reward}
In the traditional CTDE framework~\cite{CTDE}, the rewards are shared by different agents. However, in our scenario, the objectives and optimization variables are highly different. e.g., If VU $n$ is allocated to do the local computation, the rewards for energy consumption and resolution degeneration are irrelevant to $Agent_2$. Therefore, we design the rewards for $Agent_1$ and $Agent_2$ separately. Moreover, as we consider a user-centric scenario, we decompose the rewards for $Agent_1$ and $Agent_2$ among different VUs, and construct a novel algorithm in Section~\ref{sec:algorithms}.

\subsubsection{$Agent_1$ Reward for each VU $R_1^t[n]$}
The $Agent_1$ is responsible for channel access arrangement, which takes an important position in the whole system. The rewards for any VU $n$ in $Agent_1$ contain: (1) a descending reward $R_r^t[n]$ for different received frame resolutions from high to low. (2) a penalty for every transmission failure $R_f^t[n]$. (3) a weighted penalty $R_e^t[n]$ for energy consumption corresponding to VU's battery state: $\omega_e\times e_{n,l}^t(\mu_n)$. To fulfill our objective in Eq.~(\ref{eq:objectivefunction}), we give (4) a huge reward/penalty $R_{w}^t[n]$ according to the \textbf{Worst VU} at the final step: $\omega_{end} \times \left ( \min\limits_{n\in\mathcal{N}} \left[\left(\sum_{t=1}^T I_n^t\right)-\tau_{n,F}\right] \right)$. However, the sparse reward is devastating to a goal-conditioned DRL environment without taking any actions~\cite{sparsereward}. Thus, only giving a huge penalty at the final step can make the rewards ``sparse'', as the reward for FPS can not be shown to the agent, and this metric is very important in the proposed problem. Then, it will be hard to train and slow to converge. To avoid the sparse reward and accelerate training, we set an \textbf{early termination flag}. For any VU, if the frame failure times exceed the tolerant failure times, (failure times of $\text{VU}_n$ $> T-\tau_{n,F}$), we assign an (5) additional early termination penalty $R_{term}^t[n]$ according to the number of left frames: $\omega_f \times (T-t)$, and end this episode immediately. $\omega_e, \omega_{end}, \omega_f$ above are all hyper-parameters to be set in the experiments.

\subsubsection{$Agent_2$ Reward for each VU $R_2^t[n]$}
The $Agent_2$ is responsible for the downlink power allocation. We weed out some rewards from $R_1^t[n]$ that are not related to $Agent_2$. Therefore, the rewards for every VU $n$ in $Agent_2$ contain: (1) different resolution reward: $R_{r}^t[n]$, (2) transmission failure penalty $R_f^t[n]$ (3) the huge penalty for Worst VU: $R_{w}^t[n]$, and (4) the early termination (task fail) penalty $R_{term}^t[n]$.

For every reward, we narrow the range by dividing the number of VUs $N$ to ease the training.

\section{Our User-centric MADRL Approach} \label{sec:algorithms}
Our proposed User-centric Critic with Heterogeneous Actors (UCHA) structure uses the state-of-the-art Proximal Policy Optimization (PPO) algorithm as the backbone. Inspired by the effective Hybrid Reward Architecture (HRA)~\cite{HRA}, we design a user-centric Critic, which evaluates the current state-value in a more user-specific view. And considering the heterogeneous action space that incorporates both discrete and continuous actions, we create a heterogeneous Actor structure. Thus, the preliminaries, PPO (as the backbone) and HRA (as the inspiration structure) will first be introduced. We will then explain UCHA.

\subsection{\textbf{Preliminary}}
\subsubsection{Backbone: Proximal Policy Optimization (PPO)}
\textbf{Why PPO?} As we emphasize developing a \textit{user-centric} model which considers VUs' varying purpose of use and requirements and uses multiple agents with heterogeneous actions, the algorithm's policy stability and the ability for dealing with both discrete and continuous actions are essential. Therefore, the Advantage Actor-Critic structure based algorithms~\cite{A2C} that directly evaluate the V values for states instead of Q values for actions is an advisable choice. Among them, the Proximal Policy Optimization (PPO) by openAI~\cite{PPO} is an enhancement of the traditional Advantage Actor-Critic which fulfills the two requirements with better sample efficiency by using two separate policies for sampling and training, and it is more stable by applying the policy constraints.

PPO has been actively used in solving wireless communication problems and Metaverse~\cite{wfiot, MECRLsurvey}, and its prowess has been demonstrated in many scenarios, like the recent widely discussed chatbot, \mbox{ChatGPT}~\cite{ChatGPT}. Next, we will introduce the pipeline of PPO and expound on its two pivotal features: (i) Importance sampling, and (ii) Policy constraint.

\textbf{Use of importance sampling.} Importance sampling (IS) refers to using another distribution to approximate the original distribution~\cite{owen2000safe}. In order to increase the sample efficiency, PPO uses two separate policies (distributions) for training and sampling to better utilize the collected trajectories, which uses the theory of importance sampling.~\cite{PPO}. To distinguish between the two policies, we use $\pi_\theta$, $\pi_{\Bar{\theta}}$ to denote the policies for training and sampling, where $\pi$ is the policy network and $\theta, \Bar{\theta}$ are the parameters. In practice, the Temporal-Difference-1 (TD1) state-action pairs are used, then, the objective function is reformulated as:
\begin{align}
    J(\theta) &=\mathbb{E}_{(s^t,a^t)\sim\pi_{\theta}} \left [\pi_\theta(s^t,a^t) A^t \right ] \nonumber\\
    &=\mathbb{E}_{(s^t,a^t)\sim\pi_{{\Bar{\theta}}}} \left [\frac{\pi_\theta(s^t,a^t)}{\pi_{{\Bar{\theta}}}(s^t,a^t)} A^t \right ] \nonumber \\ 
    &\approx \mathbb{E}_{(s^t,a^t)\sim\pi_{{\Bar{\theta}}}} \left [\frac{\pi_\theta(a^t|s^t)}{\pi_{{\Bar{\theta}}}(a^t|s^t)} A^t \right ], \label{eq:IS}
\end{align}
$A^t$ is short for the advantage function $A(s^t,a^t)$ (a standard term in reinforcement learning to measure how much is a certain action a good or bad decision given a certain state). How to sensibly estimate the $A$ has been widely discussed. In this paper, the advantage function is estimated by the truncated version of generalized advantage estimation (GAE)~\cite{GAE}, which will be explained in ~\ref{sec:UCHA}. We assume $\pi_{\theta}(s^t) = \pi_{\Bar{\theta}}(s^t)$ here as calculating the probabilities of states occurrence is impractical~\cite{PPO}, and we use $\approx$ instead of $=$ to avoid misunderstandings.

\textbf{Add KL-divergence penalty.}
To increase the stability, we need to reduce the distance between the two distributions $\theta$ and $\Bar{\theta}$. Therefore, Trust Region Policy Optimization (TRPO)~\cite{TRPO}, which is the predecessor of PPO, uses a Kullback-Leibler (KL) divergence constraint to the objective function to limit the distance between the two distributions by directly setting $D_{KL}(\pi_{\theta}||\pi_{{\Bar{\theta}}}) \leq \varrho$, where $\varrho$ is a hyper-parameter which will be set in their experiments. Nonetheless, this constraint is imposed on every observation, and it is very hard to use in practice. Thus, PPO re-formulates its objective function into the following~\cite{PPO}:
\begin{align}
    \Delta\theta = \mathbb{E}_{(s^t,a^t)\sim\pi_{{\Bar{\theta}}}}[\triangledown f^t(\theta,A^t)],\label{eq:actorobj}
\end{align}
where
\begin{align}
    f^t(\theta, A^t)\hspace{-2pt}=\hspace{-2pt}\min\{\frac{\pi_{\theta}(a^t|s^t)}{\pi_{\Bar{\theta}}(a^t|s^t)}A^t,
    clip(\frac{\pi_{\theta}(a^t|s^t)}{\pi_{\Bar{\theta}}(a^t|s^t)}, 1\hspace{-2pt}-\hspace{-2pt}\epsilon, 1\hspace{-2pt}+\hspace{-2pt}\epsilon)A^t\}.
\end{align}

\textbf{Critic loss.} Considering the particularity of our Critic structure, we separate the Critic and Actor networks, instead of using the shared layers like the implementation in the famous Stable Baselines3 (SB3)~\cite{SB3}. The loss function is as follows:
\begin{align}
    L(\phi) = (V_\phi(s^t)-V^t_{target})^2, \label{eq:criticloss}
\end{align}
where $V^t_{target} = A^{GAE}+ V_{\phi'}(s^{t})$. The $V$ is the Critic (value) network, and $\phi$ is the parameters. Then, $V_{\phi}(s)$ means the state-value~\cite{RLintro} of state $s$ generated by Critic network. Note that $\phi'$ is the parameter of the target Critic network, and it will be replaced periodically by $\phi$. This prevailing trick is to increse the stability of the target~\cite{RLintro}. Besides, the advantage function is estimated by the GAE algorithm, and we simply use $A^{GAE}$ here.

\subsubsection{Hybrid Reward Architecture (HRA)}

\begin{figure}[t]
\centering
\includegraphics[width=0.9\linewidth]{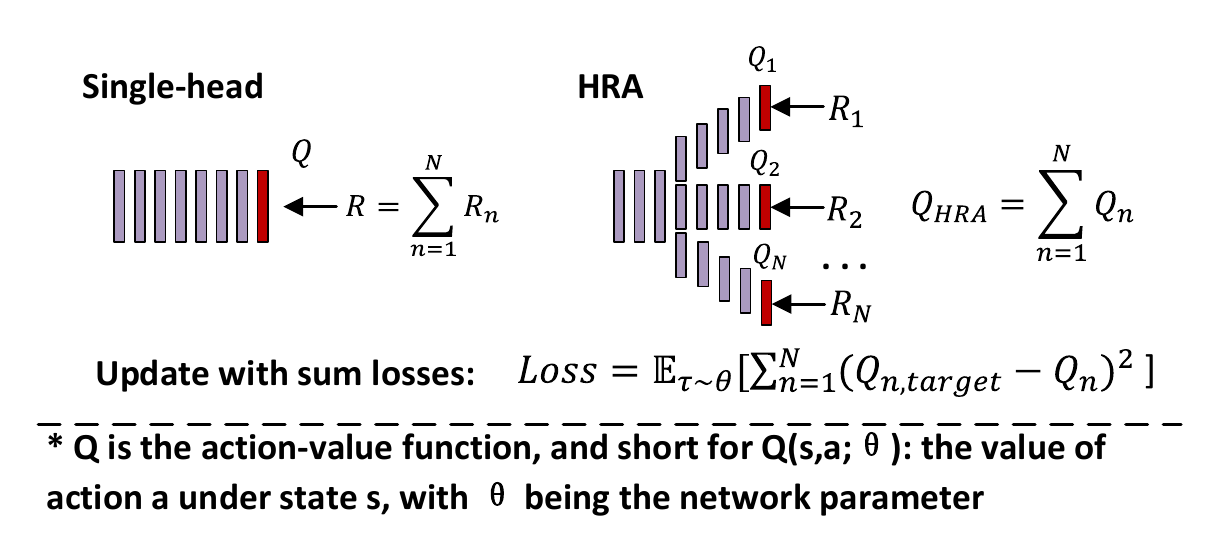}
\caption{The Hybrid Reward Architecture from~\cite{HRA}. The reward is decomposed into multiple domains, and different from the normal value network which only outputs one Q value for the action, HRA outputs multiple Q values for every domain.}
\label{fig:hra}
\end{figure}

Since there are always mixed objectives in the communication scenarios that lead to hybrid rewards reflected in reinforcement learning, we usually need to find a solution to tackle the hybrid reward problem. In this scenario, the hybrid rewards refer to the different rewards for different VUs, as we consider their distinct inherent characteristics and requirements. This issue of using RL to solve a high dimensional objective function was first studied in~\cite{HRA}. In their work, they proposed the HRA structure for Deep Q-learning (DQN) which aims to decompose rewards into different reward functions according to each objective. e.g., Decomposing rewards into objectives for energy consumption and delay minimization, and calculating two different losses accordingly. The illustration of HRA is shown in Fig.~\ref{fig:hra}. HRA can exploit domain knowledge to a much greater extent and has remarkable effects on environments with different roles and objectives, which has been demonstrated to speed up learning with better performances in various domains such as video game playing~\cite{HRAgame}, mimicry learning, complementary task~\cite{HRAmimic}, etc. However, to our best known, no one has explored this structure in depth and used it in solving communication problems. This paper re-design the normal Critic into a user-centric Critic by embedding HRA structure. Specifically, we decompose the rewards into different VUs instead of different objectives and calculate the losses of each VU accordingly. Next, we will expound on our novel and effective algorithm, User-centric Critic with Heterogeneous Actors (UCHA).

\subsection{User-centric Critic with Heterogeneous Actors (UCHA)}
\label{sec:UCHA}
In contrast to decomposing the overall reward into separate sub-goal rewards as done in HRA, we built a user-centric reward decomposition Critic, which takes in the rewards of different users and calculates the actions-values separately. In other words, we give the network a view of the value for each user, instead of merely evaluating the overall value of an action based on an overall state. Simultaneously, considering the highly different roles, action spaces, and objectives of the two agents, we design the Heterogeneous Actors structure. The two agents are updated by different rewards and advantages, which will be expounded in the following.

\begin{figure}[t]
\centering
\includegraphics[width=1\linewidth]{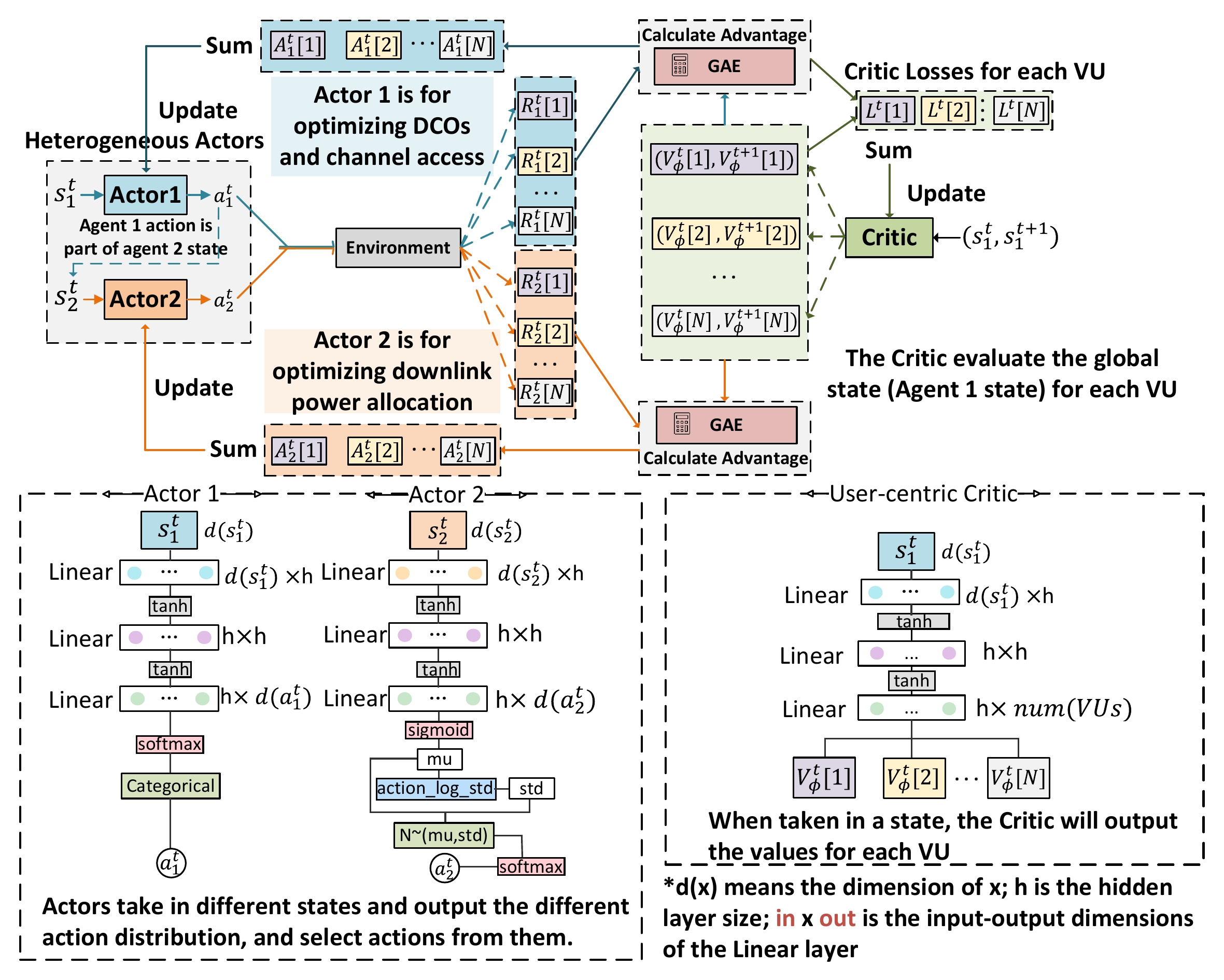}
\caption{The structure of User-centric Critic with Heterogeneous Actors (UCHA). The top of the figure is the function process and overall map of UCHA, while the bottoms are the network structures in practice (coding). The two distinct Actors take in different states and select different actions, and the Critic evaluates the global state, outputting values for each VU, separately.}
\label{fig:alg}
\end{figure}

\textbf{Function process:} In each episode, when the current transmission is accomplished with the channel access arrangement $a_1^t$ from $Agent_1$, and downlink power allocation $a_2^t$ from $Agent_2$, the environment will issue every VU's rewards $\boldsymbol{R}_1^t=\{R_1^t[1], R_1^t[2],..., R_1^t[N]\}$ for $Agent_1$, and $\boldsymbol{R}_2^t=\{R_2^t[1],R_2^t[2],...,R_2^t[N]\}$ for $Agent_2$ as feedback to different VUs. The global state $s_1^t$ and the next global state $s_1^{t+1}$ will be sent to the user-centric Critic to generate the state values for each VU. Then, the rewards $\boldsymbol{R}_1^t$ and $\boldsymbol{R}_2^t$ with state-values will be used to calculate the advantages $\boldsymbol{A}_1^t=\{A_1^t[1], A_1^t[2],..., A_1^t[N]\}$ and $\boldsymbol{A}_2^t=\{A_2^t[1], A_2^t[2],..., A_2^t[N]\}$ for $Agent_1$ and $Agent_2$. The user-centric Critic takes in the global state and advantages for calculating the Critic losses $\{L_1,L_2,...,L_N\}$for different VU, and updates with the sum losses similar to HRA. Note that the \textbf{global} state ($s_1^t$), reward ($\boldsymbol{R}_1^t$), and advantage are all from $Agent_1$, as those of $Agent_2$ are mainly a portion extracted from $Agent_1$. In terms of the Actors, we use $\boldsymbol{A}_1^t$ and $\boldsymbol{A}_2^t$ to update Actors of $Agent_1$ and $Agent_2$, separately. The above-mentioned process is illustrated in Fig. \ref{fig:alg}.

\textbf{Heterogeneous actors update function:} In equation~(\ref{eq:actorobj}) (i.e., $\Delta\theta = \mathbb{E}_{(s^t,a^t)\sim\pi_{{\Bar{\theta}}}}[\triangledown f^t(\theta,A^t)]$), we established the policy gradient for PPO Actor, and our UCHA  uses the sum-advantages of every VU for the update. Then, we have the gradient $\Delta\theta_1$, $\Delta\theta_2$ for $Agent_1$ and $Agent_2$ as:
\begin{align}
    \Delta\theta_1 = \mathbb{E}_{(s^t,a^t)\sim\pi_{{\theta}'}}[\triangledown f^t(\theta,(\sum_{n=1}^N A_1^t[n])],\\
    \Delta\theta_2 = \mathbb{E}_{(s^t,a^t)\sim\pi_{{\theta}'}}[\triangledown f^t(\theta,(\sum_{n=1}^N A_2^t[n])].\label{eq:gradient}
\end{align}
where $A_1^t[n], A_2^t[n]$ denote the advantages of different VUs for $Agent_1$ and $Agent_2$. A truncated version of generalized advantage estimation (GAE)~\cite{schulman2015high} is chosen as the advantage function:
\begin{align}
    &A_1^t[n] = \delta_1^t[n] + (\gamma\lambda)\delta_1^{t+1}[n]+...+(\gamma\lambda)^{\bar{T}-1}\delta_1^{t+\bar{T}-1}[n],\\
    &A_2^t[n] = \delta_2^t[n] + (\gamma\lambda)\delta_2^{t+1}[n]+...+
    (\gamma\lambda)^{\bar{T}-1}\delta_2^{t+\bar{T}-1}[n],
\end{align}
where
\begin{align}
    \delta_1^t[n]=R_1^t[n]+\gamma V_{\phi'}(s_1^{t+1})[n]-V_{\phi'}(s_1^t)[n],\\
    \delta_2^t[n]=R_2^t[n]+\gamma V_{\phi'}(s_1^{t+1})[n]-V_{\phi'}(s_1^t)[n].
\end{align}
$\bar{T}$ specifies the length of the given trajectory segment, $\gamma$ specifies the discount factor, and $\lambda$ denotes the GAE parameter. And these parameters will be specified in experiments. $V_{\phi'}$ is the target Critic with parameters $\phi'$, which will be periodically replaced by $\phi$. The user-centric Critic has $N$ outputs, where $V_{\phi'}(\cdot)[n]$ means the Critic-head for VU $n$, which takes in the state, and outputs the state-value for VU $n$. Note that we both use $s_1^t$, but with their separate rewards, because $s_1^t$ is the global state, and we only give the Critic an overall view of the whole task.

\textbf{User-centric Critic update function:} In this user-centric Critic, we compute the value losses for each VU separately to enable  UCHA to have a user-specific view. Similar to the update method in HRA~\cite{HRA} that uses the sum losses of different components, we use the sum losses of different VUs. Therefore, the traditional PPO-Critic loss in Eq.~(\ref{eq:criticloss}) (i.e., $L(\phi) = (V_\phi(s^t)-V^t_{target})^2$) is re-formatted into:
\begin{align}
    L(\phi) =\sum_{n=1}^N L(\phi)[n]= \sum_{n=1}^N \left (V_{\phi}(s_1^t)[n]- V_{target}^t[n]\right )^2, \label{eq:HRloss}
\end{align}
where $V_{target}[n] = A_1^t[n]+V_{\phi'}(s_1^{t})[n]$, and $L(\phi)[n]$ is the Critic loss for VU $n$.

As explained above, this user-centric Critic uses centralized training with equation~(\ref{eq:HRloss}). The computation complexity with real-time expenditure in experiments will be discussed in Section~\ref{sec:experiment}.

\begin{figure}[!t] 
    \vspace{-15pt}
        \renewcommand{\algorithmicrequire}{\textbf{Initiate:}}
        \renewcommand{\algorithmicensure}{\textbf{Output:}}
        \begin{algorithm}[H]
            \caption{\label{alg:PPO} User-centric Critic with Heterogeneous Actors}
            \begin{algorithmic}[1]
                \REQUIRE Actor 1 parameter $\theta_1$, Actor 2 parameter $\theta_2$, Hybrid critic parameter $\phi$ and target network $\phi'$, initial state $s_1^0$.
                \FOR{iteration = $1,2...$}
                    \STATE $Agent_1$ executes action according to $\pi_{\theta_1^{'}}(a_1^t|s_1^t)$
                    \STATE Get reward $R_1^t[1],R_1^t[2],...,R_1^t[N]$ and the current state of $Agent_2$: $s_2^t$
                    \STATE $Agent_2$ executes action according to $\pi_{\theta_2^{'}}(a_2^t|s_2^t)$
                    \STATE Get reward $R_2^t[1],R_2^t[2],...,R_2^t[N]$ and the next state of $Agent_1$: $s_1^{t+1}$, and $s_1^t \leftarrow s_1^{t+1}$ 
                    \STATE Sample \hspace{-2pt}\{$s^t, a^t, (R_1^t[1],...,R_1^t[N]), (R_2^t[1],...,R_2^t[N]),s^{t+1}$\} till end
                    \STATE Compute advantages \{$A_1^t[1],A_1^t[2],...,A_1^t[N]$\} for $Agent_1$, \{$A_2^t[1],A_2^t[2],...,A_2^t[N]$\} for $Agent_2$, and target values\{$V_{target}^t[1],V_{target}^t[2],...,V_{target}^t[N]$\}
                    \FOR{$k$ = $1,2,...,K$}
                        \STATE Shuffle the data's order, set batch size $bs$
                        \FOR{$j$=$0,1,...,\frac{\textbf{Trajectory length}}{bs}-1$}
                            \STATE Compute gradients for Actor 1 and Actor 2 using Eq.~(\ref{eq:gradient}). 
                            \STATE Update Actors separately by gradient ascent
                            \STATE Compute Value losses for each VU 
                            \STATE Update Critic with MSE loss using Eq.~(\ref{eq:HRloss})
                        \ENDFOR
                        \STATE Assign target network $\phi' \leftarrow \phi$ every $C$ steps
                    \ENDFOR
                \ENDFOR
            \end{algorithmic}
        \end{algorithm}
\end{figure}

\section{Experiments} \label{sec-Experiments}
In this section, we conduct extensive experiments to compare UCHA with the baselines, highlighting the remarkable performance achieved by it. We will first introduce our baseline algorithms and metrics, and then present our results with detailed analysis.
\label{sec:experiment}

\subsection{\textbf{Baselines}}
As we introduce a novel UCHA algorithm with unique actor and critic structures, we weed out the key points of UCHA one by one, and design the following baseline algorithms:
\begin{itemize}
    \item \textbf{Independent PPO (IPPO)}. The most intuitive way of using RL in such a cooperative interactive environment is to implement two independent RL agents interacting with each other. We implemented two independent PPO with user-centric Critics for optimizing the channel access and downlink power allocation. They both use a normal and separate Actor and Critic and update with their own states, actions, and rewards. \textbf{This baseline is to examine the effect of the heterogeneous Actors structure in UCHA}.
    \item \textbf{Heterogeneous-Actor PPO (HAPPO)}. We also implement an HAPPO structure that is similar to UCHA. The HAPPO does not apply the user-centric architecture, but merely uses one normal Critic that outputs a single value for the whole global state. Therefore, in HAPPO, we use the sum rewards of each Agent as the single reward. i.e., $Agent_1$ uses rewards $\boldsymbol{R_1^t}=\sum_{n=1}^N R_1^t[n]$, and $Agent_2$ uses $\boldsymbol{R_2^t}=\sum_{n=1}^N R_2^t[n]$. \textbf{This baseline is to testify the effect of the User-centric structure of the Critic}.
    \item \textbf{Random}. The two random Agents select actions randomly, which represents the system performance if no optimization strategy is performed. \textbf{The random policy serves as a naive baseline to show the results if no optimization has been conducted.}
\end{itemize}

\subsection{Metrics}
We introduce a set of metrics to evaluate the effectiveness of our proposed methods.

\begin{itemize}
    \item \textbf{Worst VU frames}. We define the Worst VU frames, i.e., $\min\limits_{n\in\mathcal{N}}(\sum_{t=1}^TI_n^t-\tau_{n,F})$ in Eq.(\ref{eq:objectivefunction}). The ``Worst VU'' means the VU has the biggest gap between its successfully received frames and its required FPS, and the ``Worst VU frames'' refers to this gap.
    \item \textbf{Successful FPS}. The number of successful frames among total $T$ frames determines the FPS of the VR frames and hence fluidity of the Metaverse VR experience.
    \item \textbf{The sum local energy consumption}. We first illustrate the sum energy consumption during training, and then investigate it for different VUs in detail.
    \item \textbf{The received frame resolutions}. The received frame resolutions of different VUs will be shown to testify the comprehensive view of our proposed methods.

\end{itemize}

\subsection{Channel Attenuation}

When $z_n^t=m$, the channel attenuation becomes a very important variable for the downlink rate. In this paper, the channel attenuation is simulated as $h_{n,m}^t = \sqrt{\beta_n^t}g_{n,m}^t$. The Rician fading is used as the small-scale fading, $g_{n,m}^t = \sqrt{\frac{K}{K+1}}\bar{g}_{n,m}^t + \sqrt{\frac{1}{K+1}}\tilde{g}_{n,m}^t$, where $\bar{g}_{n,m}^t$ stands for the Line-Of-Sight (LOS) component, and $\tilde{g}_{n,m}^t$ means the Non-LOS (NLOS) component that follows the standard complex normal distribution $\mathcal{C}\mathcal{N}(0,1)$. The large-scale fading $\beta_n^t = \beta_0 (L_n)^{-\alpha}$, and $L_n$ represents the distance between the $n$th VU and server. $\beta_0$ denotes the channel attenuation at the reference distance $L_0 = 1$ m. The path-loss exponent $\alpha$ is simulated as $2$, and the Rician factor $K$ is simulated as $3$. Note that as the duration of each time slot is too short, we don't consider the geographical movements of VUs in each time step.

\begin{figure*}[t]
\centering
\subfigtopskip=1pt
\subfigbottomskip=1pt

\subfigure[Training reward with 6 VUs.]{
\begin{minipage}[t]{0.32\linewidth}
\centering
\includegraphics[width=1\linewidth]{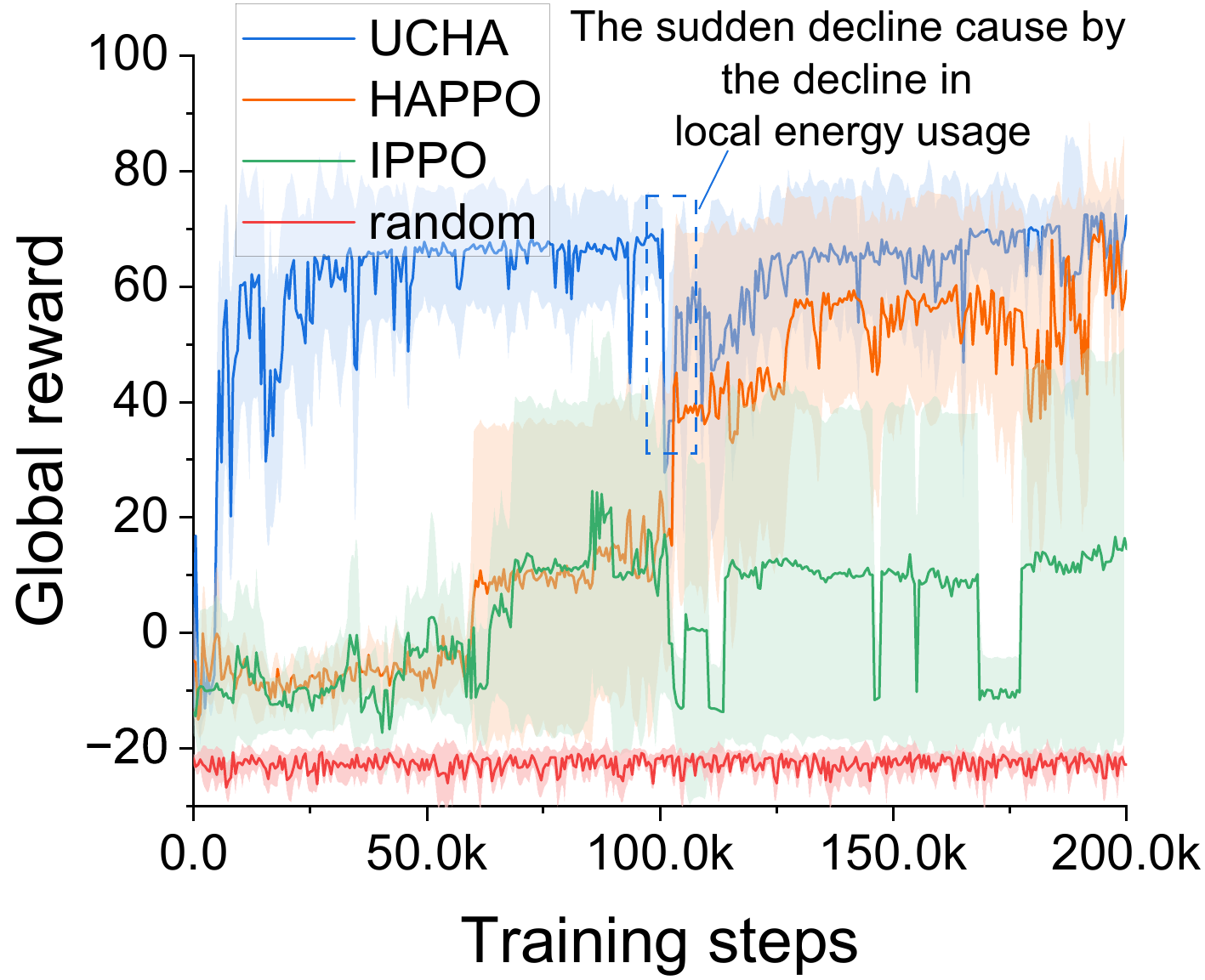}
\label{fig:36reward}
\end{minipage}
}%
\subfigure[Energy consumption with 6 VUs.]{
\begin{minipage}[t]{0.32\linewidth}
\centering
\includegraphics[width=1\linewidth]{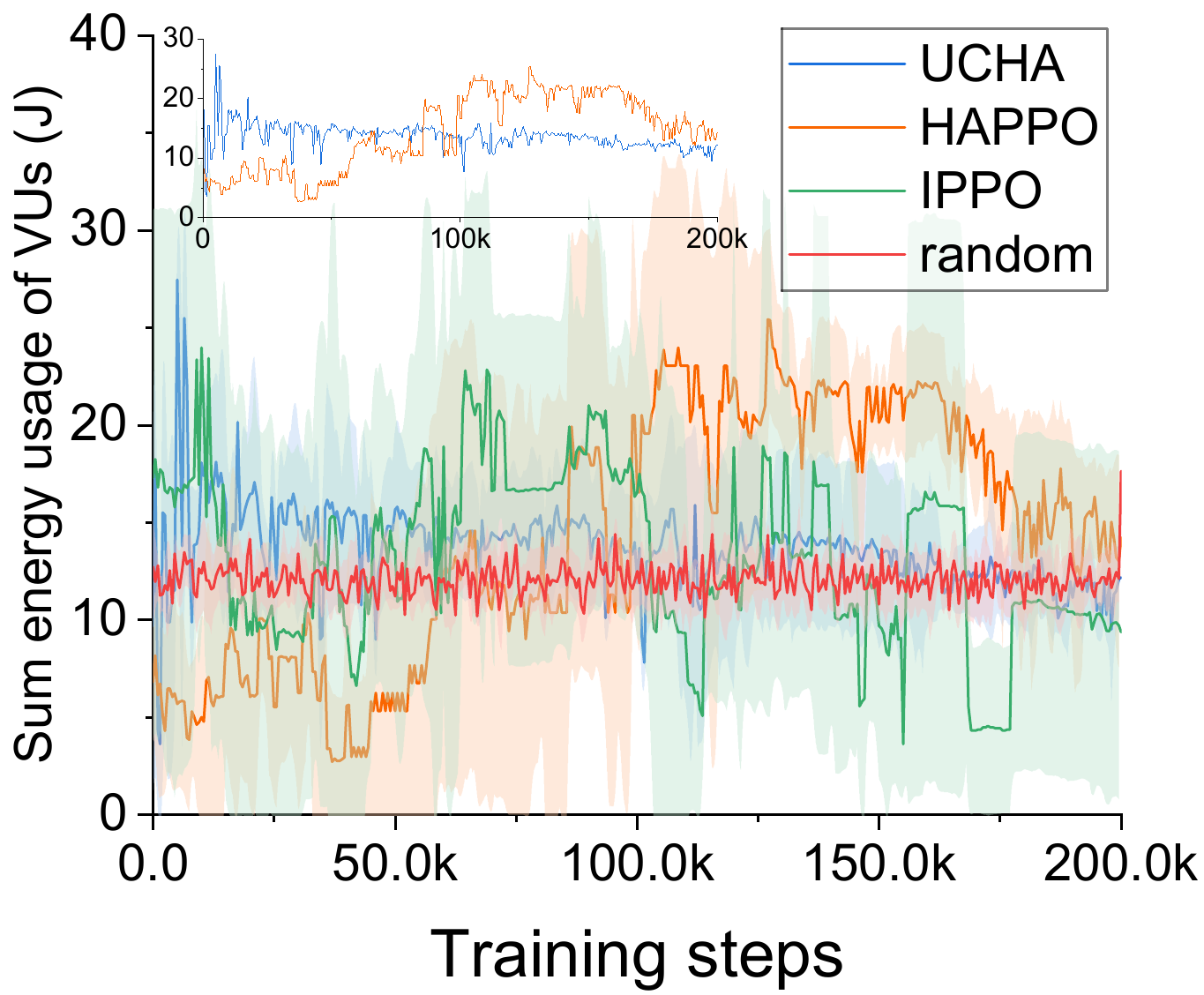}
\label{fig:36energy}
\end{minipage}%
}%
\subfigure[Worst VU frames with 6 VUs.]{
\begin{minipage}[t]{0.32\linewidth}
\centering
\includegraphics[width=1\linewidth]{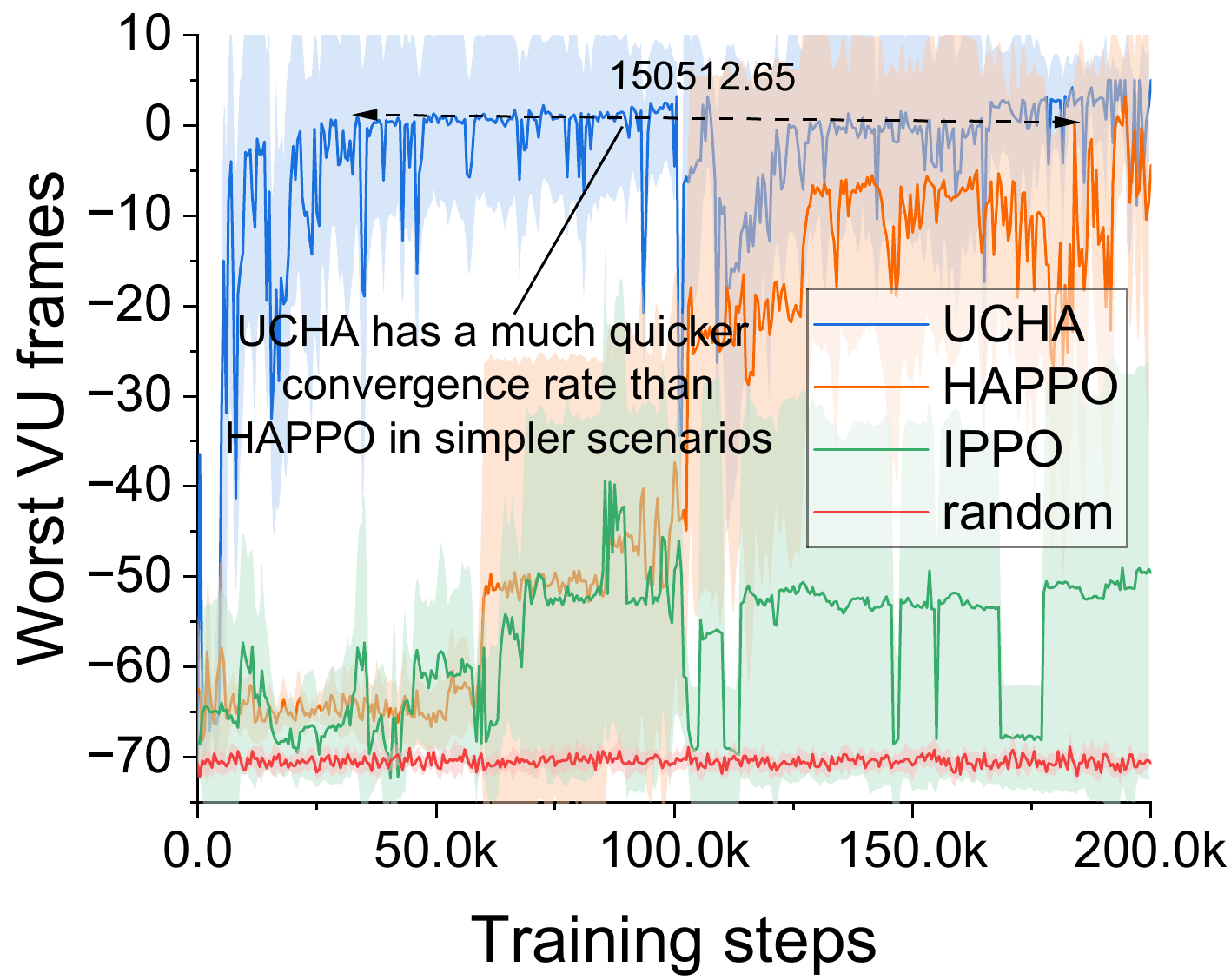}
\label{fig:36worstvu}
\end{minipage}
}%

\subfigure[Training reward with 8 VUs.]{
\begin{minipage}[t]{0.32\linewidth}
\centering
\includegraphics[width=1\linewidth]{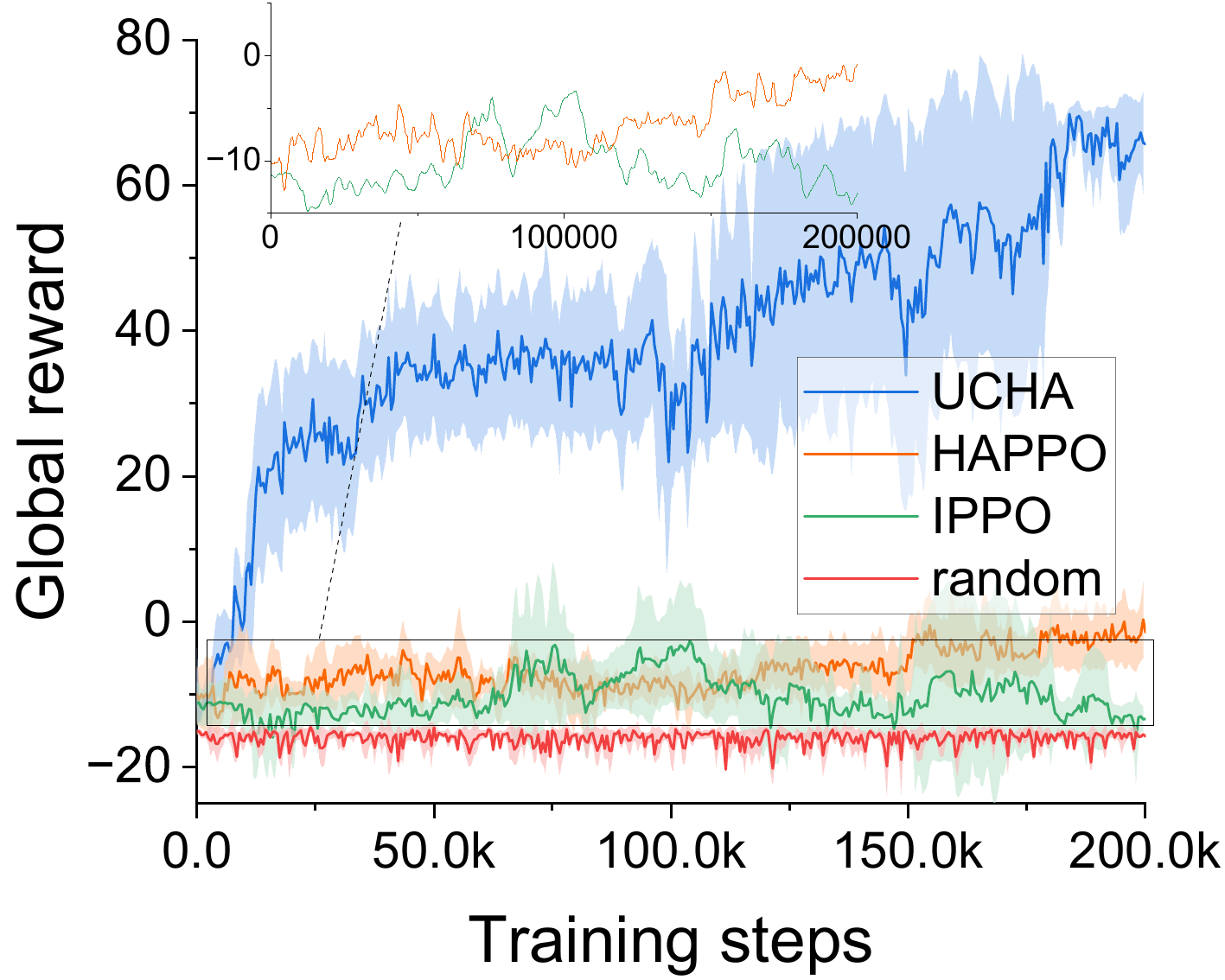}
\label{fig:38reward}
\end{minipage}
}%
\subfigure[Energy consumption with 8 VUs.]{
\begin{minipage}[t]{0.32\linewidth}
\centering
\includegraphics[width=1\linewidth]{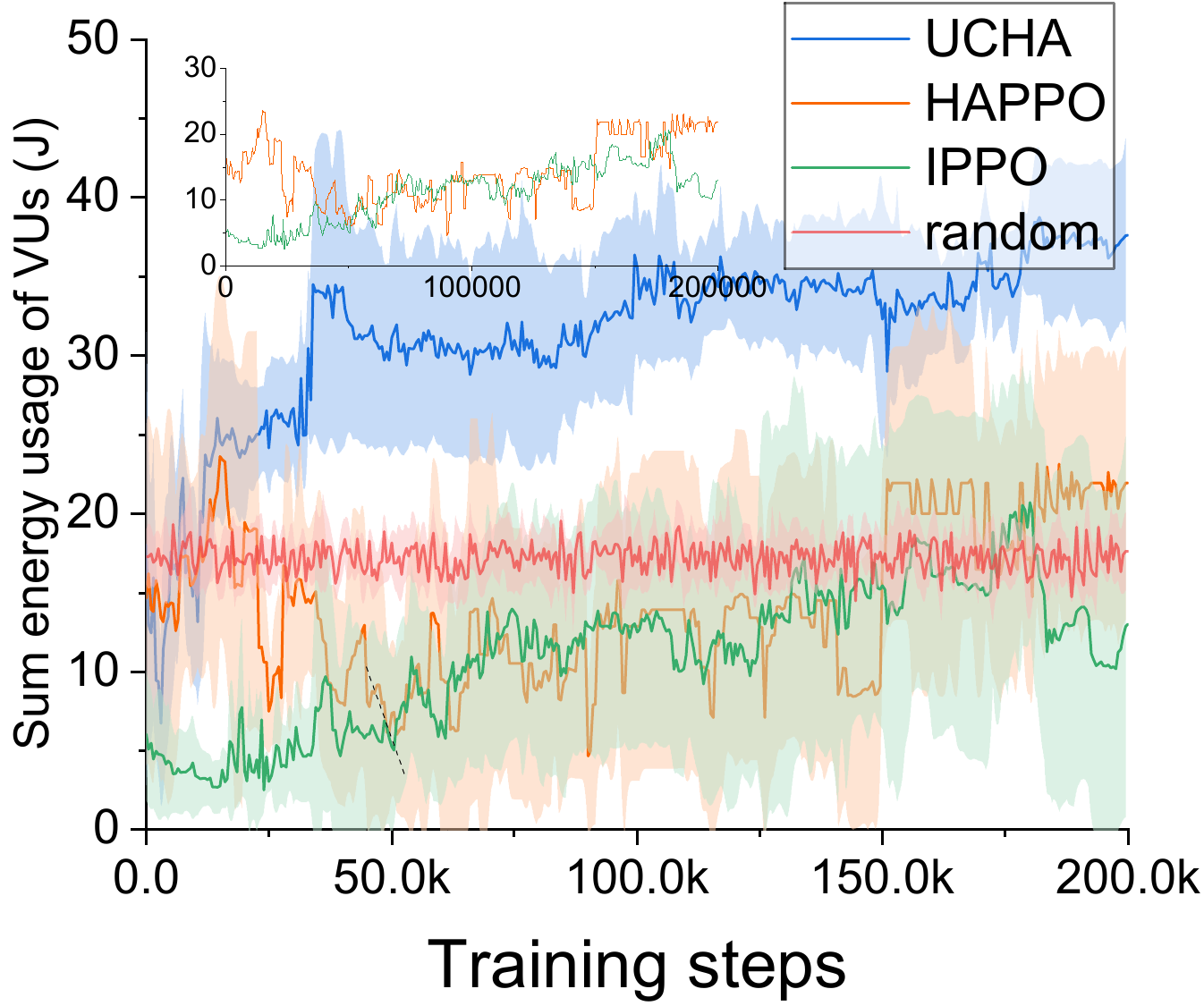}
\label{fig:38energy}
\end{minipage}%
}%
\subfigure[Worst VU frames with 8 VUs.]{
\begin{minipage}[t]{0.32\linewidth}
\centering
\includegraphics[width=1\linewidth]{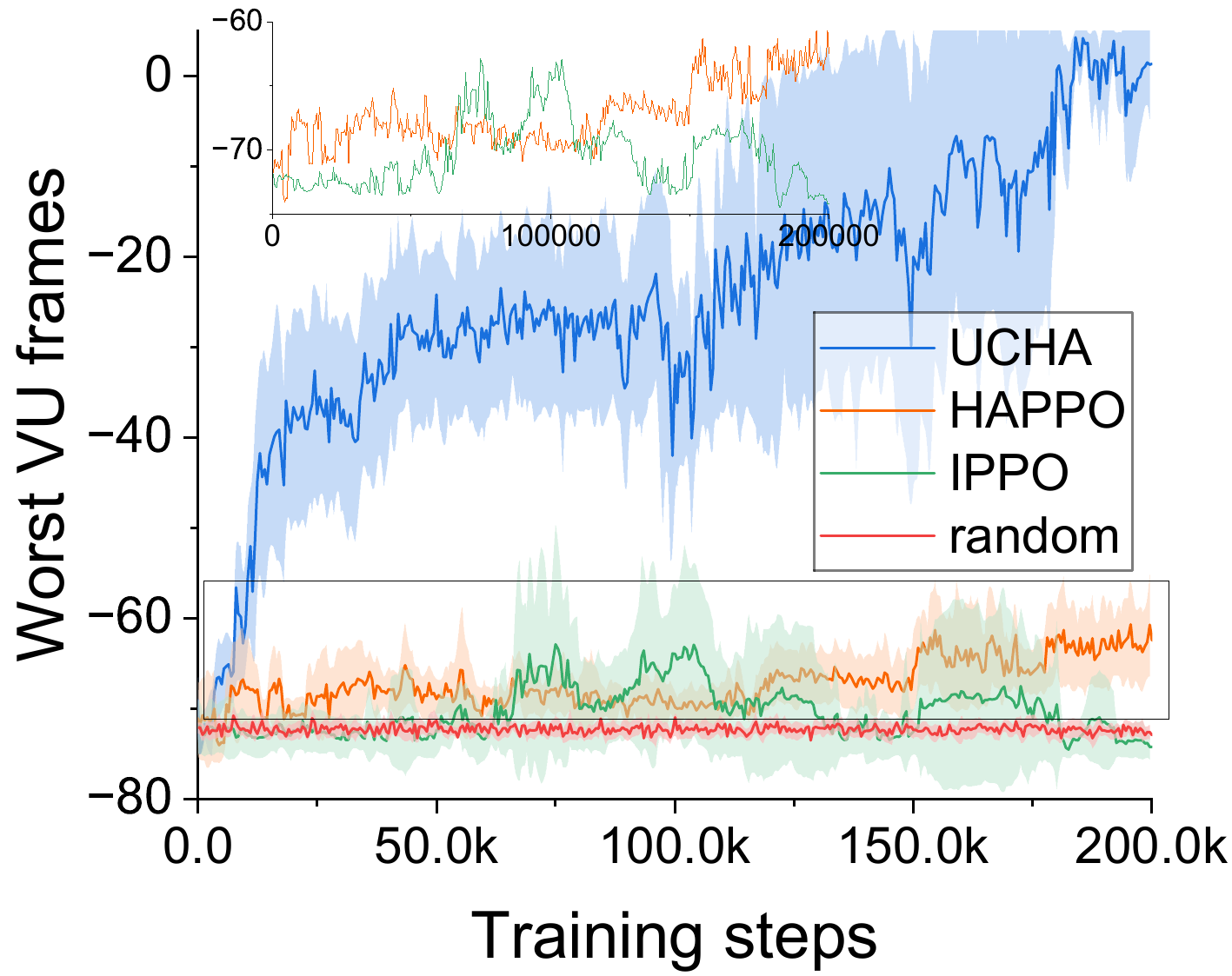}
\label{fig:38worstvu}
\end{minipage}
}%

\caption{Train-time model performances with 6 and 8 VUs (3 channels). In the simpler configuration (6 VUs), UCHA has a much quicker convergence rate than other baselines, and in the more complicated configuration (8 VUs), UCHA obtains much more remarkable performance in all aspects. }
\label{fig:train}
\end{figure*}

\subsection{Numerical Setting}
Consider a $30\times30$ $m^2$ indoor space where multiple VUs are distributed uniformly across the space. We set the number of channels to be $3$ in each experiment configuration, and the number of VUs to be from $5$ to $8$ across the different experiment configurations. To simplify the simulation, we use the most common standard of the resolutions but ignore the various horizontal FoVs and aspect ratios for different VR devices. We set the resolutions of one frame can be 1440p ($2560\times1440$, which is known as Quad HD), 1080p ($1920\times1080$, which is known as Full HD), 720p ($1280\times720$, which is known as HD), and below (considered as insufficient resolution). Each pixel is stored in 16 bits~\cite{oculus} and the factor of compression is selected in the uniformly random distribution $[300,600]$~\cite{compression}. And each frame consists of two images for two eyes. Therefore, the data size of each frame can be $D_n^t\in\{\frac{2560\times1440\times16\times2}{compression},\frac{1920\times1080\times16\times2}{compression}, \frac{1280\times720\times16\times2}{compression}\}$. The maximum flashed rate, $T$ frames in one second is taken to be 90, which is considered a comfortable rate for VR~\cite{VRsurvey} applications. The bandwidth of each channel is set to $10\times180$ kHZ. The required successful frame transmission count $\tau_{n,F}$ is uniformly selected from $[60,90]$, which is higher than the acceptable of $60$~\cite{VRsurvey}. The maximum powers of the server are $100$ Watt. The server computation frequency is $10$ GHz and the computation capabilities of each VU are selected uniformly from $[0.3, 0.9]$ GHz. For all experiments, we use a single NVIDIA GTX 2080 Ti and $2\times 10^5$ training steps, and the evaluation interval is set to be $500$ training steps. As there are several random variables in our environment (e.g., channel attenuation, compression rates), all experiments are conducted under \textbf{ten different global random seeds}, and the error bands are drawn to better illustrate the model performances.

\subsection{Train-time result analysis}
Reasonable and comprehensive experiments are of vital importance in such a complicated scenario. In this section, we will illustrate and analyze some evaluated metrics completely during training. For brevity, we show the performances of different models in two experimental configurations: the less complicated scenario with 6 VUs and the more complicated one with 8 VUs. And the overall results of every scenario are shown in Table~\ref{table:results}. We also analyze the computational complexity, and then the value losses of each VU are illustrated to give a more ``user-specific'' analysis. 
Note that to better evaluate the different metrics, we do not apply early termination if the task fails in the evaluation stage.

\begin{figure*}[t]
\centering
\includegraphics[width=0.85\linewidth]{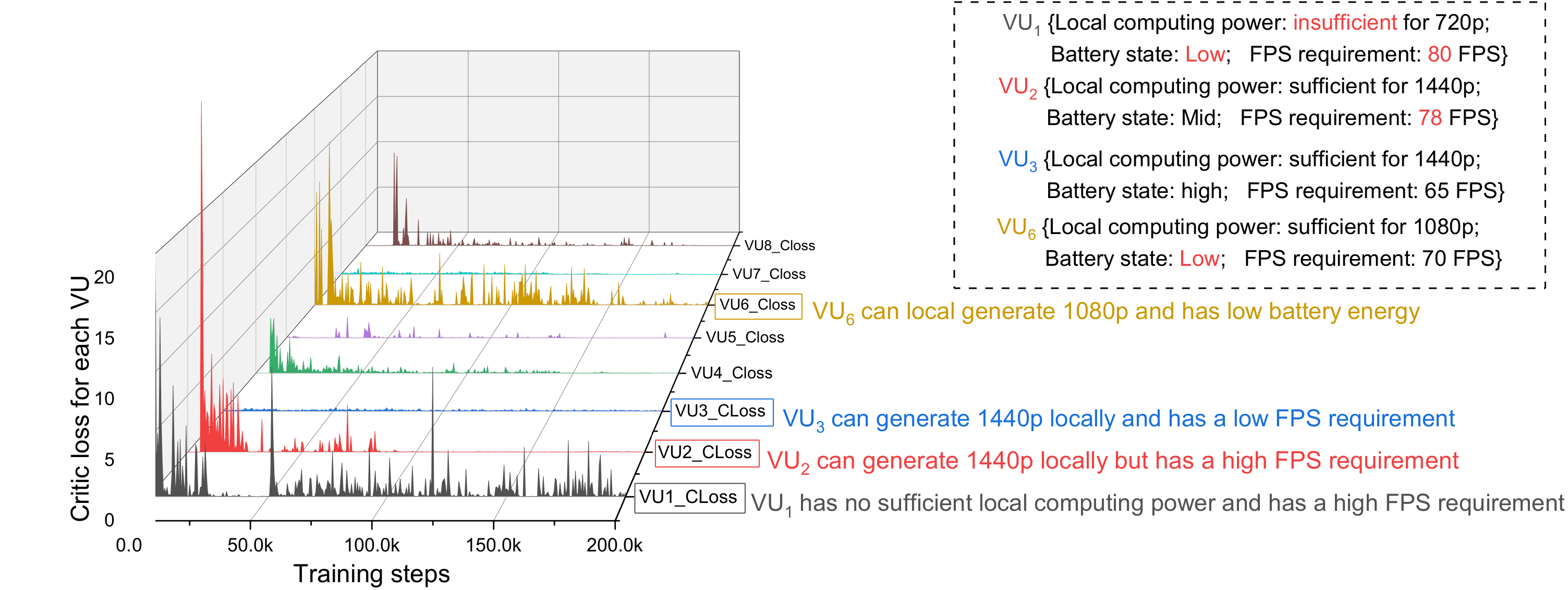}
\caption{The Critic losses (Value losses) for each VU when using UCHA in 3 channels 8 VUs scenario.}
\label{fig:vloss}
\end{figure*}

\subsubsection{Train-time model performance in ``$3\text{ Channels, }6\text{ VUs}$'', ``$3\text{ Channels, }8\text{ VUs}$'' configurations} The training reward, sum local energy consumption, and worst VU frames show an overall upward trend as training progresses. When pitted against these metrics, UCHA performed the best out of the tested baseline algorithms in both configurations. 

In the experimental setting with $6$ VUs, although HAPPO (without the user-centric structure in Critic) is able to attain similar peak rewards in the end training stages, UCHA converges in about one-fifth of training steps taken for HAPPO to achieve convergence, as shown in Fig.~\ref{fig:36worstvu}. As for the energy usage shown in Fig.~\ref{fig:36energy}, both UCHA and HAPPO increase the energy consumption (i.e., do more local generation) first, and decrease it in the following steps, while the training rewards both increase steadily. Apparently, both UCHA and HAPPO learned to decrease the congestion degree in each channel by allocating some VUs for local generation successfully. They initially tried to decrease congestion degrees to fulfill the requirements of the FPS by allocating more VUs for local generation, and then saving the local battery energy by decreasing the local generation times gradually. There is a sudden drop in the training reward of UCHA at about $100$k steps (Fig.~\ref{fig:36reward}), which is assumed to be the sharp decrease in energy usage (local computation times). This means that UCHA tried to further save energy usage by a fast decline in the computation times, but too many VUs sharing the limited channel and downlink power resources can be devastating to the overall transmission process. On balance, UCHA and HAPPO both succeed in lowering local energy usage to a similar level as if no optimizations are conducted (random policy). As a matter of fact, lowering energy usage to this random policy level is impressive. Because intuitively, we need to sacrifice more local energy to fulfill the task, and the reward and worst VU frames they obtained are extremely higher than the random policy.

Different from UCHA and HAPPO, IPPO fails to find an acceptable solution even in this simpler scenario, but we can still observe from Fig.~\ref{fig:36reward} that the reward is increasing during training. IPPO has the worst stability across different global random seeds, which is reflected in the huge error bands and the sudden drops even in the late training stages. The above observations signify that the Agents in IPPO fail to work cooperatively and they don't achieve good arrangements on channel access and downlink power compared to UCHA and HAPPO. However, IPPO still has a good performance in the simplest $3\text{ Channels, }5\text{ VUs}$ scenario. (Table~\ref{table:results})

The results of the experiment with $8$ VUs show that UCHA is extremely superior to HAPPO and IPPO in almost all aspects. Compared to the simpler configuration with $6$ VUs, UCHA used many more steps to reach the peak reward, as shown in Fig.~\ref{fig:38reward}. Moreover, in this scenario, the local energy consumption (local generation times) is increasing all the time for UCHA, which is different from the observations in the $3\text{ Channels, }6\text{ VUs}$ configuration. It is reasonable because the scenario with more VUs needs more local computation arrangements to avoid unacceptable interference. And the rationality of doing so is reflected in the steadily increasing reward (Fig.~\ref{fig:38reward}) and the slower rise of the energy usage (Fig.~\ref{fig:38energy}). The complete results are shown in Table~\ref{table:results}. We can observe that UCHA obtains the best performance for almost every metric under every scenario. This demonstrates that decomposing the reward and using sum-losses which provides a user-centric view for the RL agent, is a good approach to tackling the formulated problem. 

\begin{table}[t]
\centering
\caption{Overall results}
\label{table:parameter}
\vspace{-0.1mm}
\scalebox{0.8}{
    \begin{tabular}{cccccc}
        \hline
        Scenario & \makecell{Energy \\usage\\ (in Joules)} & \makecell{Worst VU} & Reward & \makecell{Train \\step-time \\(ms)} & \makecell{Execution \\step-time \\(ms)}\\ \hline
        \multicolumn{6}{c}{UCHA} \\ \hline
        $3\text{ Channels, }5\text{ VUs}$ & $11.78$ & $9.97$ & $80.04$ & $19$ & $0.95$\\
        $3\text{ Channels, }6\text{ VUs}$ & $10.84$ & $4.76$ & $69.72$ & $18$ & $0.94$\\
        $3\text{ Channels, }7\text{ VUs}$ & $23.67$ & $5.43$ & $62.78$ & $20$ & $0.94$\\
        $3\text{ Channels, }8\text{ VUs}$ & $37.25$ & $2.56$ & $65.6$ & $22$ & $0.98$\\ \hline
        \multicolumn{6}{c}{HAPPO} \\ \hline
        $3\text{ Channels, }5\text{ VUs}$ & $14.46$ & $4.33$ & $75.46$ & $15$ & $0.92$\\
        $3\text{ Channels, }6\text{ VUs}$ & $13.37$ & $-1.53$ & $64.98$ & $15$ & $0.91$\\
        $3\text{ Channels, }7\text{ VUs}$ & $9.88$  & $-43.76$ & $16.10$ & $16$ & $0.90$\\
        $3\text{ Channels, }8\text{ VUs}$ & $21.96$ & $-60.67$ & $-2.83$ & $18$ & $0.96$\\ \hline
        \multicolumn{6}{c}{IPPO} \\ \hline
        $3\text{ Channels, }5\text{ VUs}$ & $10.76$ & $1.32$ & $72.22$ & $20$ & $0.55$\\
        $3\text{ Channels, }6\text{ VUs}$ & $9.41$ & $-49.67$ & $14.46$ & $27$ & $0.53$\\
        $3\text{ Channels, }7\text{ VUs}$ & $10.96$ & $-66.34$ & $-16.81$ & $32$ & $0.55$\\
        $3\text{ Channels, }8\text{ VUs}$ & $12.23$ & $-73.83$ & $-14.00$ & $56$ & $0.57$\\ \hline
    \end{tabular}
}
\label{table:results}
\end{table}

\begin{figure*}[t]
\centering
\subfigtopskip=1pt
\subfigbottomskip=1pt
\subfigure[The received frame resolutions of each VU. The table on the left shows the VU inherent characteristics (i.e., Max resolution when local computing, Battery state, and FPS demand), and the right bar chart compares the received resolutions frames of each VU in one second when using different algorithms.]{
\begin{minipage}[t]{0.85\linewidth}
\centering
\includegraphics[width=1\linewidth]{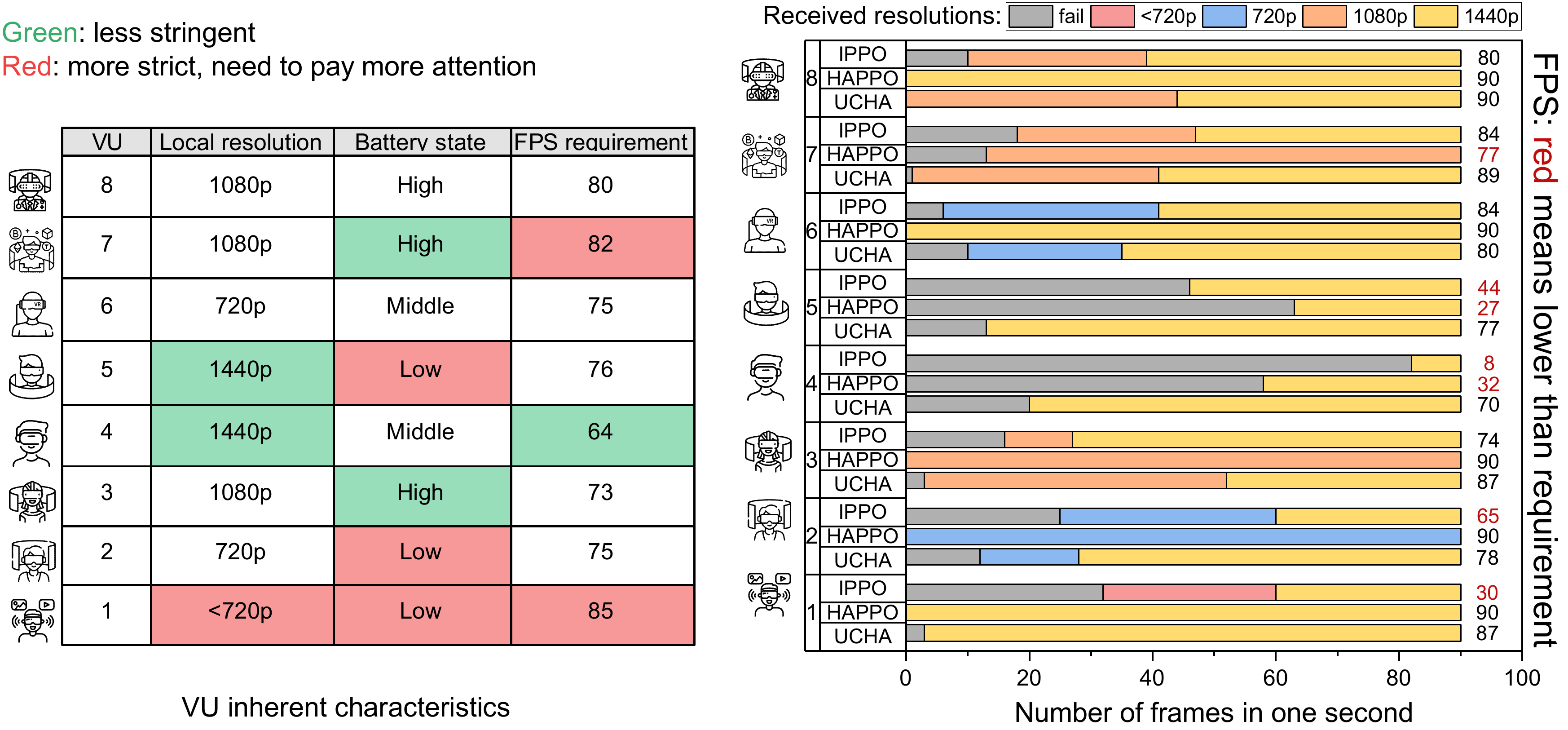}
\label{fig:resolution}
\end{minipage}
}%

\subfigure[The local computation times and energy usage. The color of the VU index denotes the battery states.]{
\begin{minipage}[t]{0.85\linewidth}
\centering
\includegraphics[width=1\linewidth]{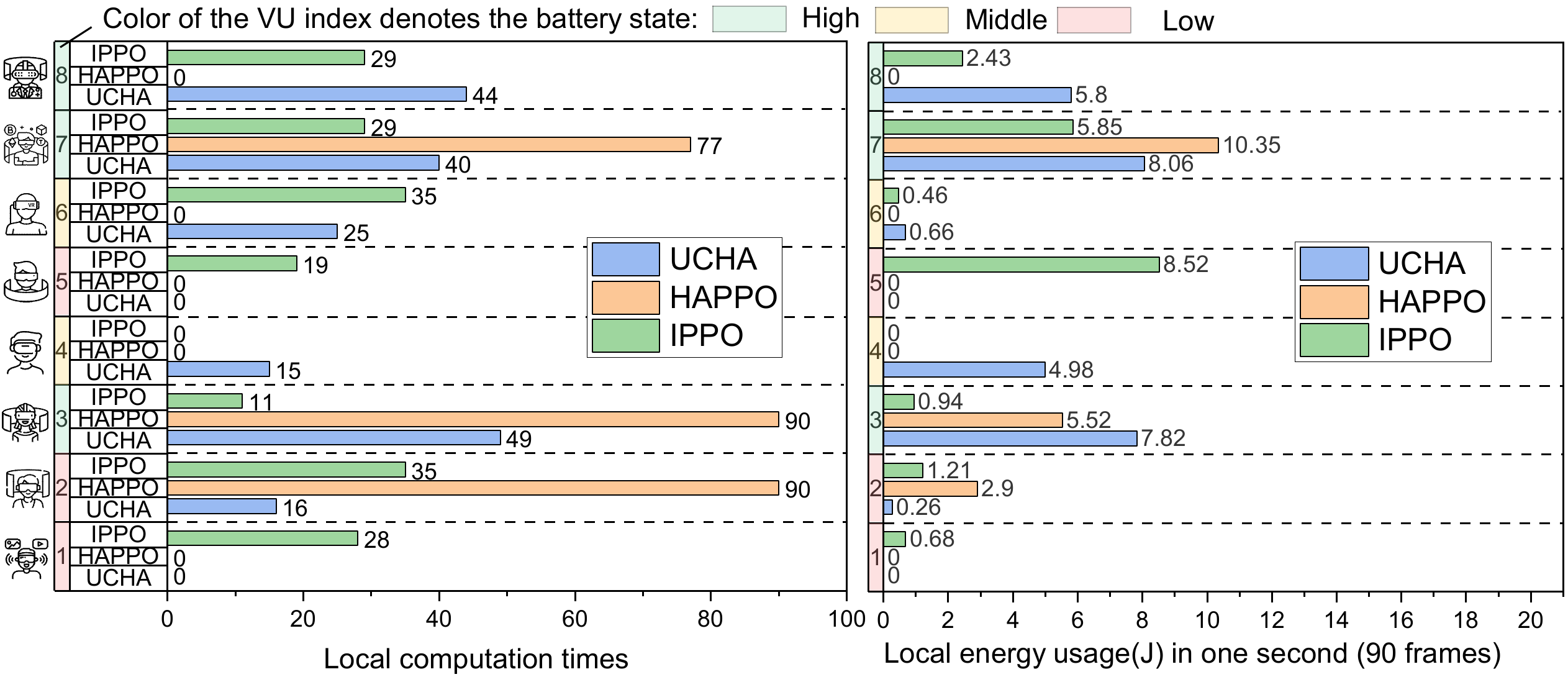}
\label{fig:userenergy}
\end{minipage}%
}%
\caption{Detailed user-specific evaluation.}
\label{fig:train}
\end{figure*}

Then we illustrate the value losses for each VU (calculated by the user-centric Critic) to show the convergence of our proposed UCHA in Fig.~\ref{fig:vloss}. Overall, all losses of VUs are decreasing, which shows the convergence of the hybrid Critic even with eight branches (one branch for one VU). However, the range and the decline patterns of the losses are all different. Here, we list four representative VUs for detailed discussion. In this setting, $\text{VU}_1$ is the toughest one for optimizing as it has no sufficient local computing power for a high resolution (i.e., higher than 720p), which is doomed to frame loss, and it has a high FPS requirement (80 FPS). Therefore, $\text{VU}_1$ must be arranged carefully, or it can cause huge losses. $\text{VU}_2$ has a high loss initially, as it has a high requirement of FPS. However, this can be easily tackled by allocating it with more local computation as $\text{VU}_2$ has sufficient local computing power for 1440p resolution. $\text{VU}_6$ is in a more involved situation. It can generate 1080p frames locally and has low battery energy. Thus, if it is arranged to do local computing, although there is no frame loss, it will be assigned energy consumption and resolution degeneration penalties. On the contrary, $\text{VU}_3$ can generate 1440p frames locally with high frame failure tolerance, and its loss is always low.

\subsection{Computational complexity}
We analyze the computational complexity and show the exact time by the figure. We use $K^l_{A}, K^l_{A'}, K^l_{C}$ to denote the number of neurons in layer $l$ of Actor1, Actor2, and the user-centric Critic. And $(\mathcal{A}_0, {\mathcal{A}'_0}, {\mathcal{C}_0})$, $(L_{A}, L_{A'}, L_C)$ be the size of the input layers (proportional to the state dimension shown in Fig.~\ref{fig:alg}), and the number of training layers of the three parts. Each training step contains two Actor training and one Critic training, and considering the mini-batch size $B$ in the training stage, we have the complexity in one training step as $O(B(\mathcal{A}_0K^1_{A}+\sum_{l=1}^{L_{A}-1}K^l_{A}K^{l+1}_{A} + \mathcal{A}'_0K^1_{A'}+\sum_{l=1}^{L_{A'}-1}K^l_{A'}K^{l+1}_{A'} + \mathcal{C}_0K^1_{C}+\sum_{l=1}^{L_{C}-1}K^l_{C}K^{l+1}_{C}))$. And according to~\cite{complexity}, the computational complexity depends on the total number of convergence steps to the optimal policy. In practice, the network training can be performed offline for a finite number of steps at a centralized-powerful unit (such as the server). We use Table~\ref{table:results} as an intuitive illustration for the time of a single training and execution (in ms) step. 

\subsection{Evaluating performance of the user-specific effect} 
In this section, we load the trained models of UCHA, HAPPO, and IPPO, evaluate them in the same evaluation environment. Noted that to better compare different algorithms,  we weed out the early termination constraint (i.e., if any VU's frame loss times reach the tolerant limit, this episode ends immediately) in evaluation, which means all evaluations finish the whole $90$ frames transmissions. 

Fig.~\ref{fig:resolution} presents the frame resolutions of each VU, in which the left table is the inherent characteristics of each VU (i.e., Max resolution when local computing, battery state, and FPS requirement), and the right bar chart compares the resolutions of each VU with different algorithms. And Fig.~\ref{fig:userenergy} further shows the local computation times and local energy usage of each VU. To simplify the illustration, we only set battery states as Low, Middle, and High, and a fixed local max resolution for each VU in this evaluation. Noted that the ``fail'' in the bar chart means this frame is allocated to be processed on the server, but not transmitted to VU in time, and the resolution less than 720p is also deemed as a failure. 

To begin with, only UCHA is able to fulfill the FPS requirements of different VUs, and it does obtain a ``user-specific'' ability. For those with less local computing power (e.g., $\text{VU}_1$, $\text{VU}_2$, $\text{VU}_6$) and with low batter energy (e.g.,$\text{VU}_5$), UCHA avoid allocating them to do many local computations. We can observe from Fig.~\ref{fig:userenergy} that UCHA decides to allocate $\text{VU}_1$ and $\text{VU}_5$ to do remote computing every time, as $\text{VU}_1$ with a high FPS requirement doesn't have the sufficient local computing power for 720p (i.e., the lowest resolution deemed as successful), and $\text{VU}_5$ has a low battery and a high local max resolution. Therefore, $\text{VU}_5$ can consume more energy with the high local resolution if it is allocated to computing locally. On the contrary, $\text{VU}_3$,$\text{VU}_7$,$\text{VU}_8$ are arranged to do more local computation by UCHA, as they have high battery energy and good local computation capabilities, which can spare the channel and downlink power resources to other VUs. Furthermore, we notice that $\text{VU}_4$ receives the least frames successfully, but this is reasonable, as $\text{VU}_4$ is in a less FPS-aware scenario and has the lowest FPS requirement, and at some steps, we need to allocate more resources to other VUs with higher requirements. The above remarkable performances demonstrate that UCHA has a ``user-centric'' ability.

Different from UCHA, HAPPO and IPPO fail to do well in such a user-centric task, and IPPO performs worse than HAPPO. We can observe from the local computation arrangement in Fig.~\ref{fig:userenergy} that HAPPO (without the user-centric Critic) learns a very extreme policy. $\text{VU}_2$, $\text{VU}_3$, and $\text{VU}_7$ are allocated to do local computation almost every time, as $\text{VU}_3$ and $\text{VU}_7$ have high battery energy which leads to relatively low local computing penalty. Although $\text{VU}_2$ has a low battery, it only generates 720p resolution locally, and the energy usage is lower. In terms of IPPO, the disorderly and unreasonable allocations signify that the agents in it are non-cooperative and do not achieve an overall good channel and power selection.

\section{Conclusion}
\label{conclude}
In this paper, we study a user-centric multi-user VR for the Metaverse over wireless networks. Users with varying requirements are considered, and a novel user-centric DRL algorithm called UCHA is designed to tackle the studied problem. Extensive experimental results show that our UCHA has the quickest convergence speed and achieves the highest reward than other algorithms, and UCHA has successfully gained the user-specific view. We envision our work to motivate more research on calibrating deep reinforcement learning for research problems of the Metaverse.

\section*{Acknowledgement}
This research is supported by the Singapore Ministry of Education Academic Research Fund under Grant Tier 1 RG90/22, RG97/20, Grant Tier 1 RG24/20 and Grant Tier 2 MOE2019-T2-1-176; and by the NTU-Wallenberg AI, Autonomous Systems and Software Program (WASP) Joint Project.

{
\renewcommand{\refname}{~\\[-25pt]References}



}

\appendices

\section{Implementation details}
\label{appendix:implementation}
We provide the hyper-parameters for the reward setting as a reference:
\begin{itemize}
    \item Descending rewards for different resolutions $R_r^t[n]$: 0 for [$0$, $720$p), 2 for [$720$p, $1080$p), 3 for [$1080$p, $2$k), 5 for [$2$k, $+\infty$].
    \item Transmission failure penalty $R_f^t[n]$: $-1$.
    \item Weight in the energy consumption penalty $R_e^t[n]$: $\omega_e=-0.5$.
    \item Weight in worst VU penalty $R_w^t[n]$: $\omega_{end}=-10$.
    \item Weight in early termination penalty $R_{term}^t[n]$: $\omega_f=-10$.
\end{itemize}

For the neural network hyper-parameters settings, the adaptive moment estimation (Adam)~\cite{Adam} is selected as the optimizer. The discount factor $\gamma$ for Actor1 and Actor2 are $0.99$ and $0.9$, respectively, and GAE factor $\lambda$ is fixed at $0.95$. The batch size is set to $64$, and the hidden layer widths are all simulated as $64$. The learning rates for Actors and the Critic are $2\times10^{-4}$.

\begin{IEEEbiography}
[{\includegraphics[width=1in,height=1.25in,clip,keepaspectratio]{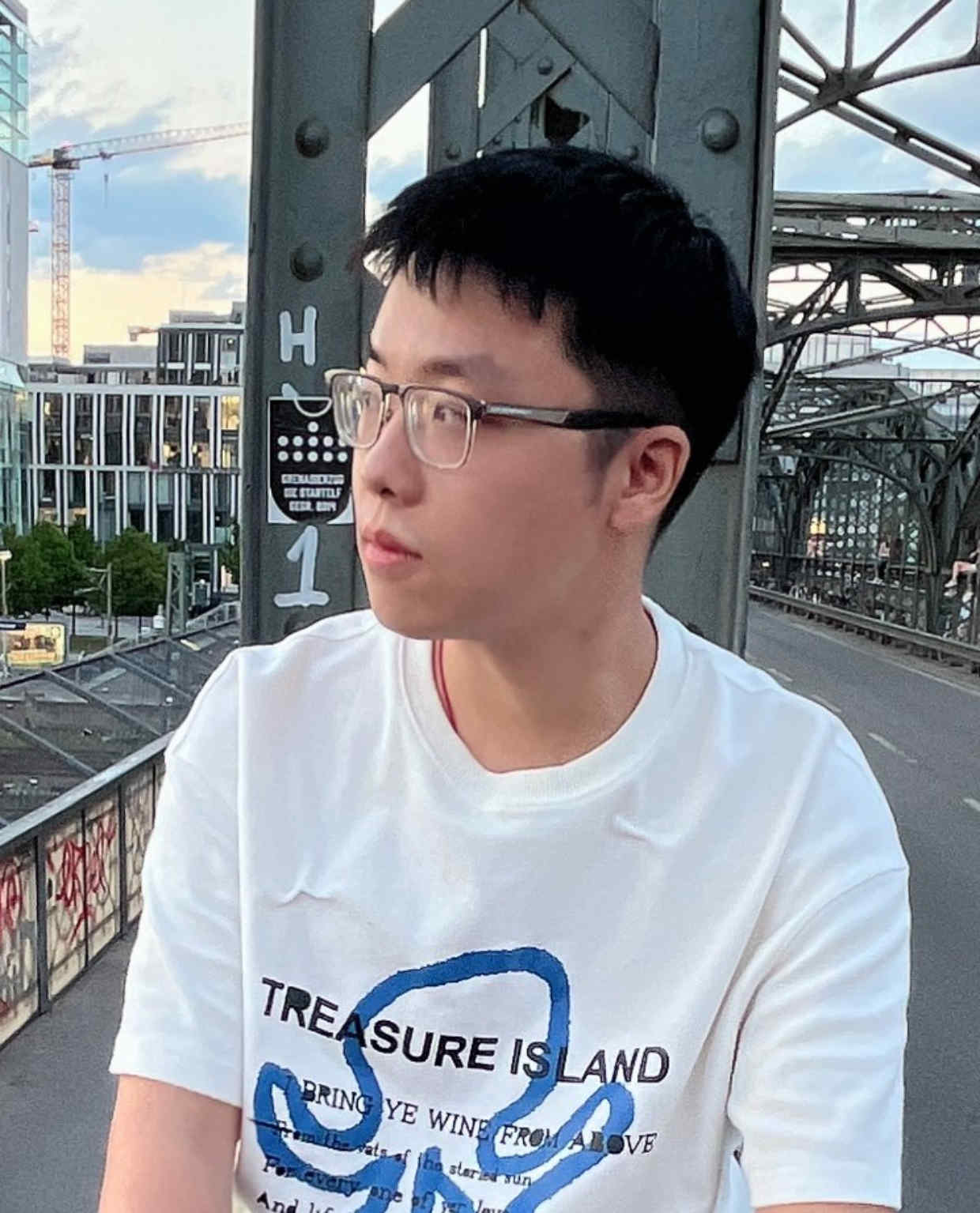}}]
{Wenhan Yu}
(Student Member, IEEE) received his B.S. degree in Computer Science and Technology from Sichuan University, Sichuan, China in 2021. He is currently pursuing a Ph.D. degree in the School of Computer Science and Engineering, Nanyang Technological University (NTU), Singapore. His research interests cover wireless communications, deep reinforcement learning, optimization, and Metaverse.
\end{IEEEbiography}
\begin{IEEEbiography}
[{\includegraphics[width=1in,height=1.25in,clip,keepaspectratio]{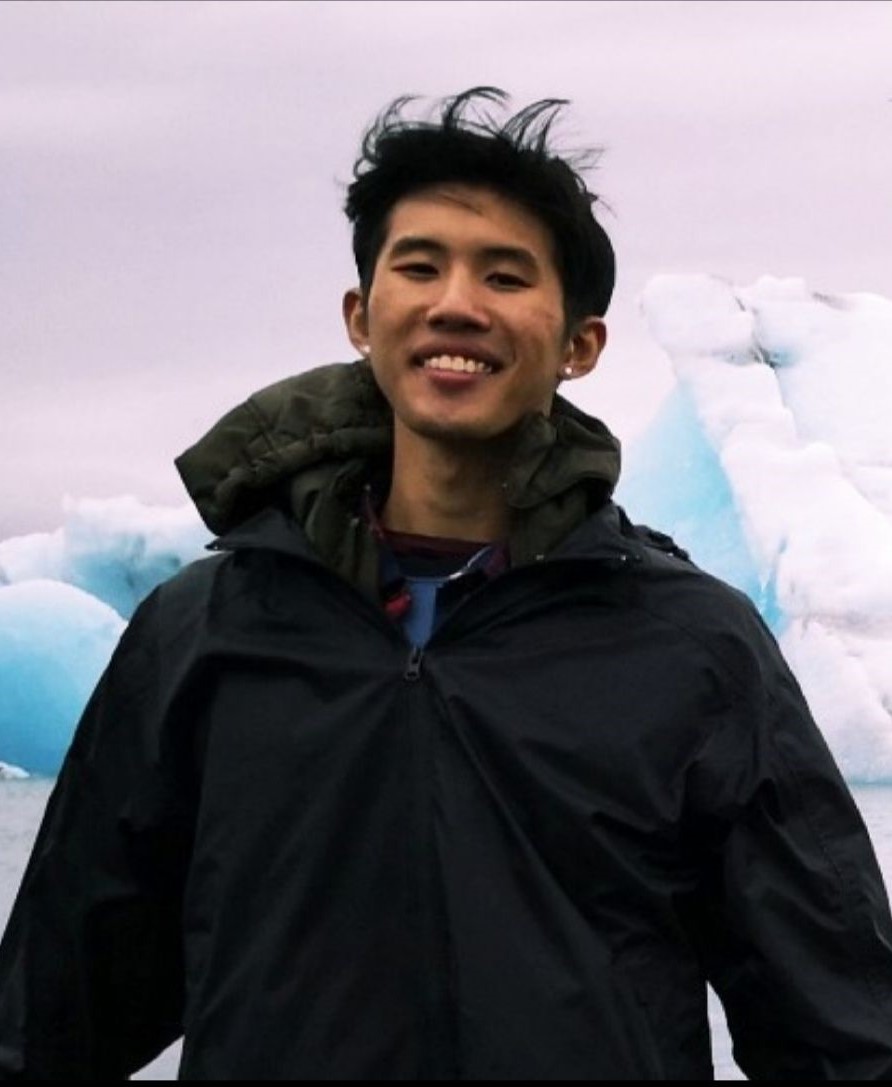}}]
{Terence Jie Chua} received his B.S. degree from Nanyang Technological University, Singapore. He is currently pursuing a Ph.D. degree in the School of Computer Science and Engineering, Nanyang Technological University (NTU), Singapore. His research interests cover wireless communications, adversarial machine learning, deep reinforcement learning, optimization, and the Metaverse.
\end{IEEEbiography}
\begin{IEEEbiography}
[{\includegraphics[width=1in,height=1.25in,clip,keepaspectratio]{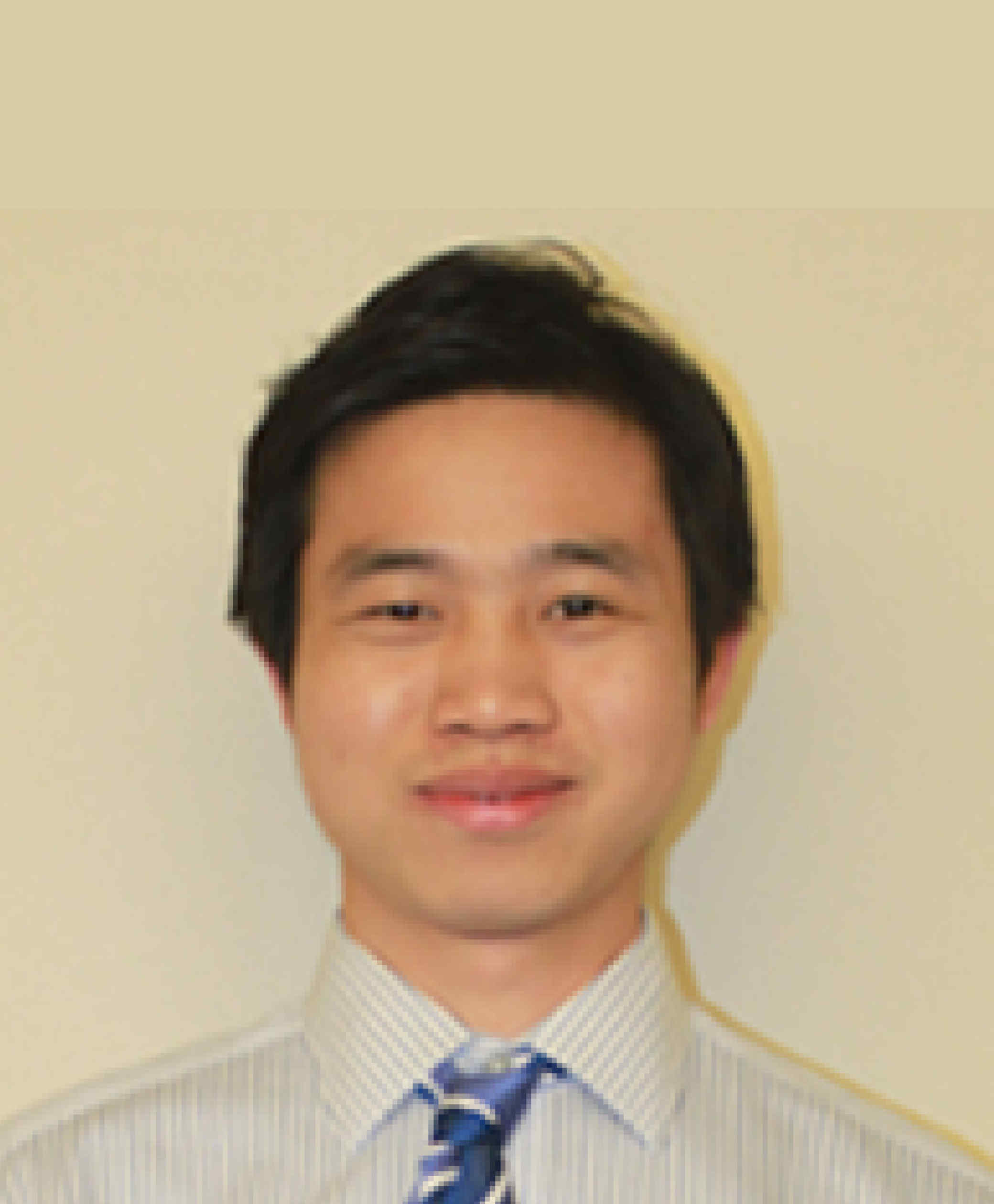}}]
{Jun Zhao} 
(Member, IEEE) received the bachelor’s degree from Shanghai Jiao Tong University, China, in July 2010, and the joint Ph.D. degree in electrical and computer engineering from Carnegie Mellon University (CMU), USA, in May 2015, (advisors: Virgil Gligor and Osman Yagan; collaborator: Adrian Perrig), affiliating with CMU’s renowned CyLab Security \& Privacy Institute. He is currently an Assistant Professor with the School of Computer Science and Engineering (SCSE), Nanyang Technological University (NTU), Singapore. Before joining NTU, he was a Post-Doctoral Researcher under the supervision of Xiaokui Xiao and then as a Faculty Member, he was a PostDoctoral Researcher at Arizona State University as an Arizona Computing Post-Doctoral Researcher Best Practices Fellow (advisors: Junshan Zhang and Vincent Poor).
\end{IEEEbiography}

\end{document}